\newif\ifmethodsformat

\methodsformattrue      % Methods-paper formatting

\ifmethodsformat

    \documentclass[9pt,twoside]{pnas-new-2019-test}
    \usepackage{metzler-preprint}

    \usepackage{newtxmath}
    \usepackage{stfloats} % allows table*/figure* at bottom of page
    
    \usepackage{cite}
    \usepackage{tabularx}
    \usepackage{array}
    \usepackage{enumitem}

    \renewcommand{\thesubsection}{\Alph{subsection}}
    \renewcommand{\thesubsubsection}{\thesubsection.\arabic{subsubsection}}

\else

    \documentclass[11pt]{article}

    \usepackage[a4paper,margin=2.15cm]{geometry}
    \usepackage[T1]{fontenc}
    \usepackage{newtxtext,newtxmath}
    \usepackage{microtype}
    \usepackage{graphicx}
    \usepackage{booktabs}
    \usepackage{tabularx}
    \usepackage{array}
    \usepackage{enumitem}
    \usepackage{float}
    \usepackage{xcolor}
    
    \usepackage{amsmath,amssymb}
    \usepackage[numbers,sort&compress]{natbib}
    \usepackage[hidelinks]{hyperref}
    \usepackage{siunitx}

    \usepackage{xltabular}
    \usepackage{makecell}

    \usepackage[font=scriptsize,labelfont=bf]{caption}

\fi

\newlist{tabitemize}{itemize}{1}
\setlist[tabitemize]{
    label=\textbullet,
    leftmargin=1.15em,
    itemsep=0.1em,
    parsep=0pt,
    topsep=0pt,
    partopsep=0pt,
    before=\vspace{-0.45\baselineskip},
    after=\vspace{-0.15\baselineskip}
}

\definecolor{qeBlue}{HTML}{2E5E88}
\definecolor{qeLight}{HTML}{EDF4F8}
\definecolor{qeGrey}{HTML}{666666}

\newcommand{\Excite}{\textbf{Excite}}
\newcommand{\Hold}{\textbf{Hold}}
\newcommand{\Release}{\textbf{Release}}
\newcommand{\Transfer}{\textbf{Transfer}}
\newcommand{\Transform}{\textbf{Transform}}

\newcommand{\perspectiveabstract}{%
A growing research community spanning quantum batteries, molecular and
excitonic energy transport, polaritonics, quantum materials and nuclear
control has begun to describe its work in terms of \emph{quantum energy} or
\emph{quantum energy science}. What connects these otherwise disparate efforts is the deliberate use of quantum states, couplings, collective effects and engineered environments to control energetic states and pathways. 
A key difference from quantum information science is that the complete quantum state does not always need to be preserved; engineered dissipation or dephasing can even improve the desired outcome. 
We organize this emerging domain through a framework that connects mechanisms, functions, capabilities, and applications. We identify five
elementary functions---\emph{Excite}, \emph{Hold},
\emph{Release}, \emph{Transfer}, and \emph{Transform}---and examine the
mechanisms that implement them. The same functions recur from the molecular
to the nuclear scale, although their relative implementation difficulty can
differ sharply. Moving toward technology, we highlight three engineering capabilities---\emph{Storage},
\emph{Routing}, and \emph{Conversion}---through the performance metrics
that matter for their applications. Stretch scenarios illustrate
what performance targets could make these capabilities
technologically consequential.
}

\ifmethodsformat

    \title{The next phase of quantum energy science:\\
    from mechanisms to applications}

    \preprintauthors{
    Florian Metzler$^{1}$,
    Matt K. Lilley$^{1}$,
    Matthew C. DeCapua$^{1}$,
    Peter L. Hagelstein$^{1}$,
    Jonah F. Messinger$^{1,2}$,
    Nicola Galvanetto$^{1,3,4}$
    }

    \preprintaffiliations{
    $^{1}$Massachusetts Institute of Technology, Cambridge, MA, USA.
    $^{2}$Cavendish Laboratory of Physics, University of Cambridge, Cambridge, UK.
    $^{3}$Univ. Grenoble Alpes, CEA, CNRS, IBS, Grenoble, France.
    $^{4}$University of Zurich, Zurich, Switzerland.
    }

    \preprintcorrespondence{}
    \preprintshortauthor{Metzler et al.}
    \preprintlabel{preprint}
    \preprintdate{October 2026}

\else

    \title{\textbf{The next phase of quantum energy science:\\
    from mechanisms to applications}}

    \author{
    Florian Metzler$^{1}$,
    Matt K. Lilley$^{1}$,
    Matthew C. DeCapua$^{1}$,
    Peter L. Hagelstein$^{1}$,
    Jonah F. Messinger$^{1,2}$,
    Nicola Galvanetto$^{1,3,4}$
    \\[0.5em]
    \small $^{1}$Massachusetts Institute of Technology, Cambridge, MA, USA\\
    \small $^{2}$Cavendish Laboratory of Physics, University of Cambridge, Cambridge, UK\\
    \small $^{3}$Univ. Grenoble Alpes, CEA, CNRS, IBS, Grenoble, France\\
    \small $^{4}$University of Zurich, Zurich, Switzerland
    }

    \date{September 2026}

\fi

\begin{document}

\ifmethodsformat

    \makepreprintopening
    {\perspectiveabstract}
    {}

\else

    \maketitle

    \begin{abstract}
    \perspectiveabstract
    \end{abstract}

    \section{Introduction: from field recognition to an engineering framework}

\fi

Quantum engineering is today most visibly associated with information. Quantum computers manipulate quantum states to process information; quantum sensors exploit state sensitivity to extract information; and quantum networks
transfer quantum states between systems. Yet the same physical systems also
carry energy. A two-level system can be viewed as a qubit whose states encode
information, or as an energetic system whose excited state stores a quantum
of energy. 
A state that is unsuitable for storing quantum information may still be useful for storing energy, and a quantum state can retain its energy even after the phase coherence involved in its preparation is lost.

In 2023, we argued that this energy-focused perspective was becoming sufficiently
coherent to warrant recognition as a counterpart to quantum information science, which we termed
\emph{quantum energy science} \cite{Metzler2023}. Tools of modern quantum
engineering---including state preparation, controllable interactions,
coherence, collective dynamics and environmental engineering---were
being deployed primarily to alter energetic outcomes. 
Our earlier article approached this emerging field mainly through the
application areas of solar energy, batteries and nuclear systems.
That application-based organization helped reveal common physics across
communities that rarely interacted. However, application labels alone
do not show which physical operations are shared or how they contribute
to a useful device. Cavity-mediated excitation transfer, for example,
can connect light-harvesting complexes (solar energy) or populate a molecular storage layer (batteries). Across research domains, experiments now demonstrate collective charging,
metastable storage, electrical extraction, cavity-mediated transfer and
direct nuclear excitation
\cite{Hu2026,Tibben2025,Hymas2026,Wu2025,Zhang2024}.
A common framework can help identify how these advances relate and
what is needed to turn them into useful devices.

\begin{figure*}[h]
    \centering
    \includegraphics[width=0.90\linewidth]{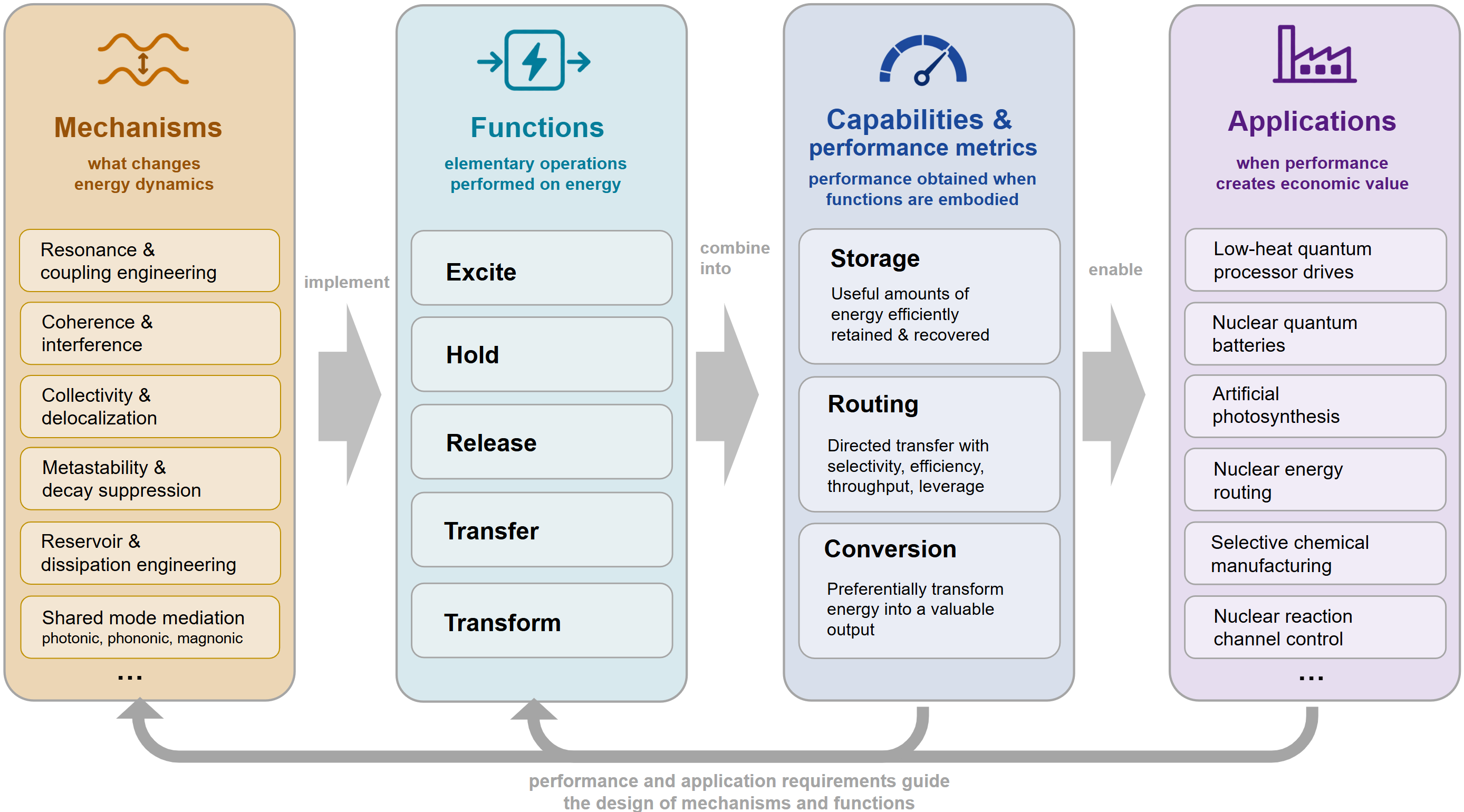}
    \caption{Maps out key relationships from mechanisms to applications in quantum energy science.
Forward arrows link mechanisms, functions, capabilities/performance metrics and applications;
lower arrows show how performance and application requirements guide
design. 
The applications
shown are the illustrative stretch scenarios in Table~\ref{tab:success}.}
    \label{fig:chain}
\end{figure*}

A sign of a maturing field is the emergence of shared conceptual
and practical vocabularies \cite{Galison1997,Kaiser2005}.
Here, we develop such a vocabulary for quantum energy science. The framework
distinguishes four linked levels, summarized in Figure~\ref{fig:chain}.
The four levels, which define the structure of this article, answer different questions. \emph{Mechanisms} describe
how quantum engineering changes energy dynamics. \emph{Functions}
describe what elementary operation is performed on energy.
\textit{Capabilities and performance metrics} describe what a combination of functions achieves and how that achievement is measured.
\emph{Applications} identify where
that performance creates economic value. The mappings are
generally many-to-many. 

We identify five elementary energetic functions: \textit{Excite}, \textit{Hold}, \textit{Release},
\textit{Transfer} and \textit{Transform}. Excite, Hold and Release describe an intuitive
sequence in which energy is injected into a particular quantum degree of freedom, is
retained, and subsequently becomes available outside a defined system boundary again. Transfer changes where
energy resides among subsystems. Transform changes the energetic carrier.
We examine three basic engineering capabilities built from these operations:
\textit{Storage}, \textit{Routing}\ and \textit{Conversion}. A storage device must load, retain and
recover useful amounts of energy; a router must deliver it to a desired destination;
and a converter must produce a useful carrier, product or material state.
Each capability is assessed through application-dependent performance
metrics. A molecular battery, for example, may use collective absorption to
implement Excite, but a Storage capability also requires Hold and
Release. Its value depends on how much energy can be recovered, at what power and how long the energy can be stored.

The functions recur across energy scales, but the constraints on their implementation
 differ. Molecular and electronic excitations in the meV--eV
range are often directly accessible to laboratory controls. For many
keV--MeV nuclear transitions, weak coupling and restrictive selection rules
make Excite and Release more demanding; the 8.4 eV nuclear transition of $^{229}$Th
is an exceptional bridge case \cite{Zhang2024}.

\section{Functions and mechanisms of quantum energy science}

\begin{figure}
    \centering
    \includegraphics[width=0.99\linewidth]{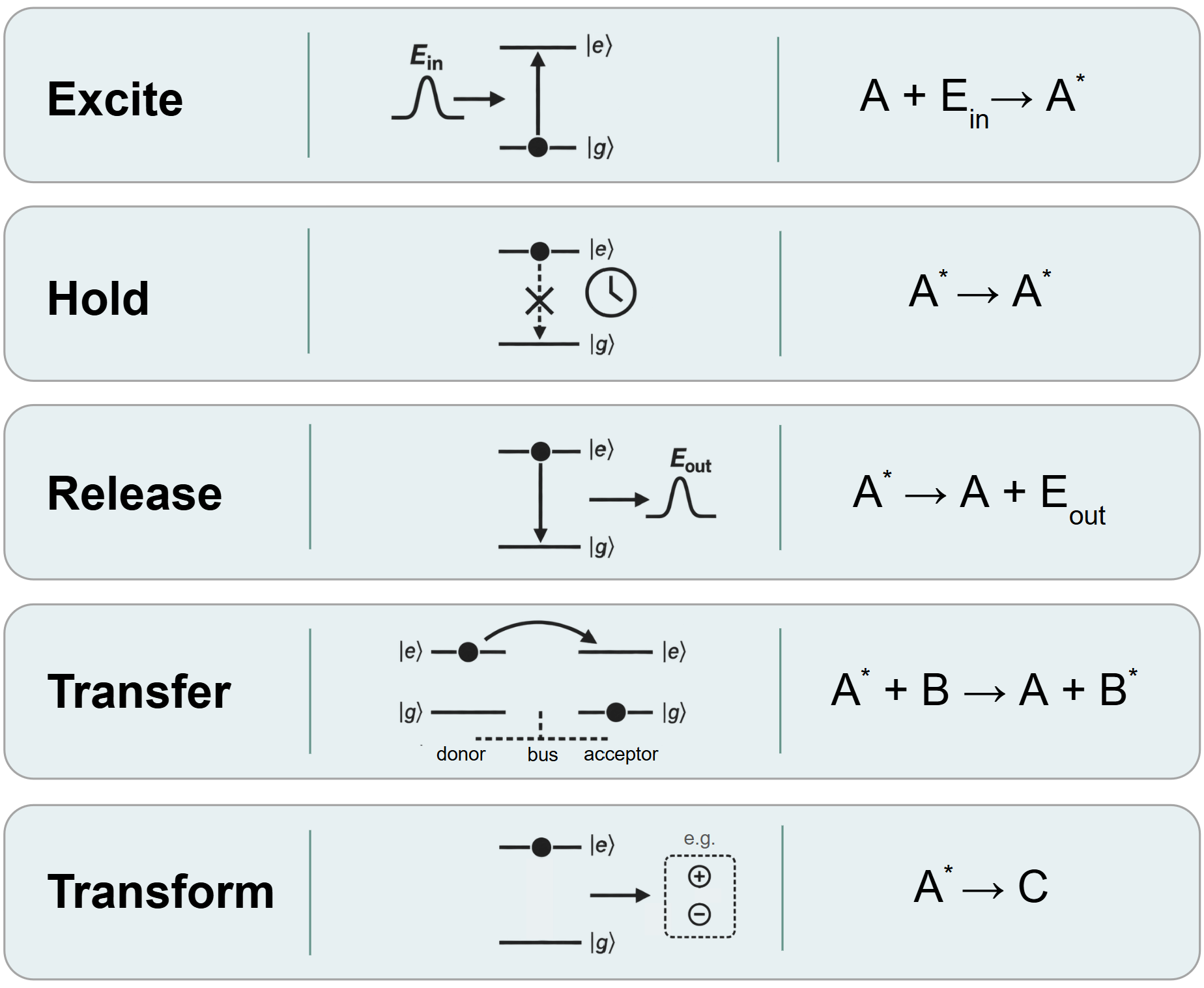}
    \caption{Five elementary energetic functions. The central sketches are
illustrative: filled dots mark initial occupation, and
\(|g\rangle\) and \(|e\rangle\) denote ground and excited states.
On the right, \(A\) and \(B\) denote subsystems and an
asterisk marks excitation. \(E_{\mathrm{in}}\) and \(E_{\mathrm{out}}\)
indicate energy entering and leaving the system.
\(C\) denotes a different energetic carrier or physical configuration,
illustrated here by charge separation.}
    \label{fig:functions}
\end{figure}

This section examines the five functions, as summarized in
Figure~\ref{fig:functions}.
Each
function is defined relative to a declared system boundary. That boundary determines whether an energy exchange is described, for example, as Release---when energy leaves the system---or as Transfer---when energy moves across coupled subsystems.

Quantum information science provides a precedent for identifying shared underlying functions in the form of the DiVincenzo criteria that are used to compare quantum computing platforms
through common operations\cite{DiVincenzo2000}. Table~\ref{tab:qis}
places essential quantum information operations loosely modeled on these criteria alongside their energetic counterparts.

\begin{table}[h]
\centering
\caption{Approximate operational correspondences between quantum information
tasks and the proposed quantum energy functions.}
\label{tab:qis}

\begingroup
\footnotesize
\linespread{1.10}\selectfont
\setlength{\tabcolsep}{3pt}

\begin{tabularx}{\linewidth}{
    >{\bfseries\raggedright\arraybackslash}p{4.8em}
    >{\raggedright\arraybackslash}X
    >{\bfseries\raggedright\arraybackslash}p{4.8em}
    >{\raggedright\arraybackslash}X
}
\toprule
\multicolumn{2}{c}{\textbf{Quantum information}} &
\multicolumn{2}{c}{\textbf{Quantum energy}} \\
\cmidrule(lr){1-2}\cmidrule(lr){3-4}

Prepare
& Initialize or write a quantum state
& \Excite
& Populate an energetic state \\
\addlinespace[0.7em]

Preserve
& Maintain coherence or fidelity
& \Hold
& Retain excitation with limited loss \\
\addlinespace[0.7em]

Output
& Measure or read out information
& \Release
& Couple stored energy to an output \\
\addlinespace[0.7em]

Transmit
& Move quantum information between systems
& \Transfer
& Move energy between coupled subsystems \\
\addlinespace[0.7em]

Process
& Apply state transformations
& \Transform
& Change the energetic carrier or physical configuration \\
\addlinespace[0.7em]

\midrule

\textit{Primary objective}
& Preserve and manipulate information-bearing states with high fidelity
& \textit{Primary objective}
& Control energy storage, delivery and conversion, with coherence where required \\

\bottomrule

\addlinespace[1.5em]

\end{tabularx}

\endgroup
\end{table}

\paragraph{Mechanisms: how the functions are implemented.}

The mechanisms highlighted in Figure~1 can be viewed as a set of physical
principles for implementing the five functions. Each changes a different aspect of the energetic dynamics: which states interact, how different probability amplitudes combine, whether constituents act independently or collectively, how readily excitation decays, how the environment redistributes energy, or which otherwise separated systems can interact.

\emph{Resonance and coupling engineering} governs which states can
exchange energy and how rapidly. Matching transition energies
can enhance exchange, while detuning can suppress it. Material
arrangement, geometry and applied fields provide ways
to adjust these interactions.

\emph{Coherence and interference} allow probability amplitudes associated with different processes leading to the same final state to reinforce or cancel one another.
Controlling their relative phases can favor some processes
and suppress others. 

\emph{Collectivity and delocalization} allow
excitation to be coherently shared across several constituents.
With suitable state preparation and couplings, their
contributions can enhance absorption, emission or energy transfer,
as in the case of superabsorption, superradiance and supertransfer
\cite{Higgins2014,Dicke1954,LloydMohseni2010}. The key distinction
is whether many constituents behave independently or whether
their probability amplitudes combine to produce collective behavior.

\emph{Metastability and decay suppression} prolong the persistence
of excited states. Selection rules or weak transition couplings
can inhibit decay as an intrinsic feature of specific states. Destructive interference can also turn states dark
or subradiant with suppressed coupling to a radiative channel. Changing the state or its couplings can then restore emission. These mechanisms control how strongly an energetic state is
connected to the channels through which its energy can escape.

\emph{Reservoir and dissipation engineering} shapes interactions
with surrounding degrees of freedom that produce relaxation and
dephasing. Controlling which environmental modes are available and
how strongly they couple can suppress unwanted decay or favor
particular relaxation routes. The environment can therefore be treated
as part of the design problem.

\emph{Shared-mode mediation} connects systems through common
electromagnetic, vibrational or spin-wave modes. When multiple systems couple
to such modes, they can mediate an indirect interaction between those systems,
potentially extending their connection beyond the range of direct
coupling. The mode's spatial profile, frequency and coupling to each
system determine which connections are effective.

The following subsections examine how these mechanisms are selected
and combined to implement the five functions.

\subsection{Excite: moving energy into a quantum system}

\textit{Excite} means deliberately increasing the energy stored in a specific quantum degree of freedom by supplying energy from outside the declared
system boundary:
\begin{equation}
    A + E_{\mathrm{in}}
    \xrightarrow{\mathrm{Excite}}
    A^* .
\end{equation}
This includes charging a quantum battery, exciting molecular vibrations
and populating nuclear excited states.
Excitation depends on how the external source couples to the states of interest, the spectral, spatial and
temporal profiles of the driving field,
and interactions within the system
and with its environment. The examples below illustrate different ways of
engineering these variables.

At molecular and electronic energy scales, excitation of target systems is often readily
accessible in principle, for example, via common optical lasers in the eV and sub-eV range. The engineering question is how rapidly and how selectively the desired states can be populated. Several distinct strategies address this problem.

One is to strengthen or reshape the coupling between the applied field and
the target transition. Mochizuki \emph{et al.} used plasmonic field enhancement to
strengthen the infrared driving interaction with surface-bound CO,
enabling successive excitation to higher vibrational states
\cite{Mochizuki2026}.
A second strategy is to make excitation cooperative rather than independent.
Coupling many constituents to a common electromagnetic mode can enhance
their collective interaction with the source, as in the superabsorptive
charging observed by Quach \emph{et al.} in an organic microcavity
\cite{Quach2022}. More generally, theory shows that collective or 
driving operations can reduce charging times relative to independent
excitation under matched driving constraints \cite{Binder2015,Gyhm2022}.
Hu \emph{et al.} demonstrated a related route in superconducting hardware
by engineering interactions that excite neighboring transmon cells in
pairs \cite{Hu2026}.
Excitation can also be improved by controlling what happens after energy
enters the system. Shastri \emph{et al.} showed theoretically that an
appropriate level of dephasing can suppress persistent energy-exchange
oscillations and drive a battery more rapidly toward a charged steady state
\cite{Shastri2025}. 
Excite can therefore be engineered both through the incident source and through the coupling architecture and the dynamics that govern the accumulation of energy in the target state.

At nuclear energies, excitation can proceed directly or through
intermediate nuclear or electronic states. Direct vacuum-ultraviolet laser excitation of the approximately
$8.4~\mathrm{eV}$ isomeric transition has been demonstrated in
$^{229}$Th-doped crystals \cite{Tiedau2024,Elwell2024,Zhang2024}. Hiraki
\emph{et al.} indirectly populated the same isomer through resonant x-ray excitation
of a higher nuclear level whose decay feeds the target state
\cite{Hiraki2024}. Kishimoto \emph{et al.} demonstrated another indirect
route in $^{197}$Au, where a nearly energy-matched electronic transition
following x-ray-induced inner-shell ionization excites the
$77.351~\mathrm{keV}$ nuclear state \cite{Kishimoto2000NEET}.

For hard-x-ray and gamma-ray nuclear transitions, the challenge is to deliver enough
radiation in the relevant energy band that couples to the desired nuclear resonance. XFELs
have directly excited hard-x-ray nuclear transitions, including the
$14.4~\mathrm{keV}$ transition of $^{57}$Fe
\cite{Chumakov2018XFEL}. MeV nuclear photoexcitation is also established using accelerator-based gamma-ray sources \cite{Loher2013NRF}, but producing substantial excited-state populations remains extremely difficult because only a small fraction of the available radiation couples resonantly to the nuclear transition. Phase-sensitive manipulation additionally requires suitable pulse coherence and temporal control \cite{Heeg2021}.
These constraints motivate engineering of both the incident field and the
nuclear environment. Lentrodt \emph{et al.} theoretically identified
thin-film cavity designs that enhance nuclear excitation by focused x-ray
pulses \cite{Lentrodt2025}, while Dahal \emph{et al.} proposed shaping
the incident x-ray waveform to enhance collective excitation of
$^{57}$Fe nuclei in a waveguide \cite{Dahal2026}. Mechanical modulation can also alter which incident frequencies couple to a nuclear transition. Nazeeri \emph{et al.} used a \(97.9~\mathrm{MHz}\) surface acoustic wave to redistribute the \(14.4~\mathrm{keV}\) Mössbauer absorption spectrum of a \(^{57}\)Fe film into sidebands \cite{Nazeeri2026SAW}. This changes the spectral accessibility of the transition.

The resulting excited population provides the starting point for Hold, Release, Transfer or Transform.

\subsection{Hold: protecting an energetic state from decay} \label{sec:hold}

\textit{Hold} means preserving excited states against unwanted decay or
leakage. Schematically,
\begin{equation}
    A^*(t_0)
    \xrightarrow{\mathrm{Hold}}
    A^*(t_1),
\end{equation}
where the repeated label $A^*$ denotes retention of energy in the quantum state. Recall that the focus is on the energy content of the quantum state and not the information content---phase coherence can decay during this interval while energy remains stored.

There are two broad approaches to Hold: one is to place the energy in a state that is intrinsically long lived or only weakly coupled to loss channels; the other is to engineer the surrounding couplings so that otherwise available escape pathways are suppressed. Molecular and electronic systems often require one or both strategies because the strong couplings that make Excite and Release easy also provide efficient routes for stored energy to escape.

A key strategy for avoiding fast decay is to move excitation from a short-lived absorber state to a longer-lived storage state. 
Tibben \emph{et al.} demonstrated
a molecular-microcavity architecture in which absorbed energy is transferred
into long-lived triplet states in a separate storage layer \cite{Tibben2025}. Stability can also be created by engineering how populated states couple to their environment.
In the molecular-microcavity model of Quach \emph{et al.}, dephasing transfers
energy from bright states into dark states, suppressing re-emission through
the cavity mode \cite{Quach2022}.
Another way to control leakage is to modify the pathways available to the
stored excitation. Chen \emph{et al.} propose a waveguide storage scheme in
which emitter--mirror interference suppresses radiative emission, while
detuning the storage and loading ensembles after excitation inhibits backflow
into the radiatively active loading ensemble \cite{Chen2026}. 
Shvyd'ko \emph{et al.} experimentally used timed magnetic switching in
$^{57}$FeBO$_3$ to alter interference among nuclear transitions and
partially suppress accelerated coherent emission after
$14.4~\mathrm{keV}$ excitation \cite{Shvydko1996}.

Long-lived nuclear states---nuclear isomers---provide an extreme example of the first strategy, in which retention is largely a property of the selected state itself. Nuclear isomerism arises when de-excitation from an excited
nuclear configuration is strongly hindered, for example by large changes in
angular momentum or by nuclear-structure effects
\cite{WalkerPodolyak2020}. Such states can retain substantial nuclear
excitation for times ranging from seconds to years or longer, motivating
longstanding interest in isomers as high-energy-density storage media
\cite{WalkerDracoulis1999,Carroll2007}.
The $^{229}$Th isomer provides a very-low-energy example. Its
approximately $8.4~\mathrm{eV}$ transition is many orders of magnitude below
typical nuclear transition energies.
In Th-doped CaF$_2$, Hiraki \emph{et al.} measured an isomeric half-life of
$447(25)~\mathrm{s}$ \cite{Hiraki2024}. In a suitable host, the
excitation can therefore be retained for minutes without continuous driving,
shifting much of the engineering challenge toward efficient preparation and
controlled access to the stored energy \cite{Zhang2024,Hiraki2024}.

\subsection{Release: opening a controlled output channel}

\textit{Release} means making energy stored inside the declared system boundary
available outside that boundary on demand:
\begin{equation}
    A^*
    \xrightarrow{\mathrm{Release}}
    A + E_{\mathrm{out}} .
\end{equation}
Release is the operational counterpart to Hold: Hold suppresses unwanted
leakage, whereas Release deliberately opens or strengthens a chosen output
channel. The output may emerge through radiative emission, work extraction,
or another controlled depletion pathway.

Two approaches to Release recur. One is to control an existing
output pathway by changing its coupling strength, timing, or extraction operation. Another is to create an alternative pathway, for example by connecting a long-lived state to a state that decays more readily.

In the first approach, the coupling to an output channel is changed dynamically. Wang \emph{et al.} propose modulating the external coupling of a microwave cavity
to coordinate energy accumulation and pulsed emission from a metastable spin
ensemble \cite{Wang2026}. In a waveguide-QED setting, Chen \emph{et al.}
similarly propose protecting excitation during storage and then restoring
resonance between storage and radiatively active ensembles to reopen a
superradiant emission channel \cite{Chen2026}.

Collective emission can also strengthen a controlled output.
Kersten \emph{et al.} stored excitation in a spin ensemble by
detuning it from a cavity, then restored resonance to enable
superradiant emission \cite{Kersten2023Triggered}. Collective interference can also make emission channel-selective rather
than simply faster. Asenjo-Garcia \emph{et al.} showed theoretically that
ordered atomic arrays can emit efficiently into a guided optical mode
while suppressing emission into unwanted free-space modes, providing a
route to preferential release through a chosen output
\cite{AsenjoGarcia2017SelectiveRadiance}. 

Control pulses can also extract stored energy as work.
Li \emph{et al.} implemented microwave-pulse sequences that
extract work from a superconducting qubit and distinguish the
contribution of coherence \cite{Li2025Ergotropy}.
Cenedese \emph{et al.} predict that monitoring light leaking from a
cavity charger can increase the recoverable work when the
subsequent work-extraction protocol is adapted to the measurement
record \cite{Cenedese2026}.

The same broad strategy, i.e., controlling an already available output
channel, extends to nuclear excitation.
R\"ohlsberger \emph{et al.}
prepared a purely superradiant state of $^{57}$Fe nuclei in a planar x-ray
cavity and observed collective decay almost two orders of magnitude faster
than the single-nucleus radiative decay
\cite{Rohlsberger2010Superradiance}. 
As discussed under Hold, 
Shvyd'ko
\emph{et al.} used magnetic switching in excited $^{57}$FeBO$_3$ to
partially suppress coherent forward emission; switching the magnetization
back restored the channel and produced a delayed x-ray burst
\cite{Shvydko1996}. Coherent control provides another handle: Heeg
\emph{et al.} used the relative phase of two x-ray pulses to switch a
$^{57}$Fe nuclear exciton between coherent enhanced excitation and coherent
enhanced emission \cite{Heeg2021}.

A harder Release problem arises when the same properties that
protect stored energy also hinder its retrieval, as with nuclear isomers. Work on the $^{229}$Th isomer illustrates how decay rate might be influenced.
In $^{229}$Th doped crystals, irradiation has accelerated its depletion by
approximately threefold with lasers \cite{Schaden2025} and up to
roughly fifty-fold with x-rays \cite{Hiraki2024}, relative to decay
without irradiation.
Release may be stimulated by opening an alternative pathway through which the stored
excitation can decay more readily.Laser-assisted coupling to electronic states in the crystal
has been proposed as a route to nuclear de-excitation
\cite{Nickerson2021}.

For higher-energy isomers, whose transitions commonly lie in the keV--MeV range, one route to Release is to populate an
intermediate nuclear state that decays more readily.
Beli\'{c} \emph{et al.} depleted $^{180\mathrm{m}}$Ta using
approximately MeV-energy photons to populate higher-energy states which subsequently decay \cite{Belic2002Ta180m}.
Ding \emph{et al.} induced depletion of $^{93\mathrm{m}}$Mo by
accelerating $^{93\mathrm{m}}$Mo ions into lead and carbon foils,
with the depletion attributed to inelastic nuclear collisions
\cite{Ding2026Mo93m}.

Other proposed approaches seek to connect the long-lived state to a more
rapidly releasing one without relying exclusively on direct high-energy
driving.
Collective excitation could increase access to otherwise weak nuclear
transitions. Dahal \emph{et al.} propose extending waveform-engineered
collective absorption to long-lived nuclear isomers and suggest that related
control principles could be used for retrieval; their quantitative
calculations concern $^{57}$Fe, so their application to isomers remains prospective \cite{Dahal2026}. Superradiant emission could similarly accelerate Release once population has been coupled
into a collectively radiative state, provided that the selection rules allow for the transition. Another possibility would be to
Transfer excitation from a long-lived isomer to a shorter-lived nuclear
acceptor and allow that acceptor to decay rapidly---the inverse of moving
short-lived excitation into a long-lived state for Hold. Other efforts to release energy from long-lived nuclear states include DARPA's \textit{Decay on Demand} program, which is investigating externally induced alpha emission from long-lived actinides to produce daughter radioisotopes for applications including nuclear power sources \cite{DARPADecayOnDemand}; the target states there are long-lived ground-state nuclides rather than metastable isomers, but the work shares the broader challenge of externally controlling decay from a long-lived nuclear state.

\subsection{Transfer: redistributing energy among coupled subsystems}

\textit{Transfer} requires at least two distinguishable subsystems:
\begin{equation}
    A^* + B
    \xrightarrow{\mathrm{Transfer}}
    A + B^* .
\end{equation}
The excitation redistributes within the system and changes its \emph{address}: the donor loses energy and an
acceptor gains it while the participating subsystems remain recognizably the
same subsystems. This definition covers resonance energy transfer, exciton
migration, state swaps between qubits, cavity- or waveguide-mediated
exchange, and nuclear excitation transfer. 
Transfer differs from Excite in that the intervention that triggers it need not supply the transferred energy: a weak control can change the couplings or resonance conditions that determine where energy goes that is already in the system.

The transfer processes considered here are nonradiative in the sense that the excitation is exchanged within the coupled system rather
than first being emitted into a freely propagating radiation field. If a
donor emits a real photon that leaves the system and that photon
is subsequently absorbed by another system, the appropriate description
is Release followed by Excite. By contrast, exchange across subsystems through
near-field coupling or a shared photonic, vibrational or magnetic mode
remains Transfer when the mediator is included within the system
boundary.

We highlight four general engineering approaches: material design can change
the direct couplings and energetic landscape connecting donor and acceptor;
shared modes can create additional indirect connections; collective or
delocalized states can make many sites participate coherently in the same
transfer; and external driving or environmental coupling can strengthen,
suppress or unblock particular transfer routes.

The first approach is illustrated by non-fullerene organic photovoltaics, where molecular organization, disorder and electronic coupling influence exciton diffusion and donor–acceptor energy transfer \cite{Wang2024NFA}. In Y-type acceptors, Guo \emph{et al.} associate faster
exciton migration with more uniform molecular packing and moderately
reduced intermolecular interactions, and find a correlation with
photovoltaic performance \cite{Guo2024ExcitonDynamics}.

A second approach is to couple donor and acceptor systems through
a shared mode. Under strong light--matter coupling, molecular
excitations and cavity photons form hybrid states, or polaritons,
that can connect distant molecules and enable transfer beyond
the range of direct intermolecular coupling \cite{Sandik2025}. Wu \emph{et al.} used enhanced exciton--exciton
annihilation as evidence for increased energy transfer between
photosynthetic light-harvesting complexes coupled to a common cavity
mode \cite{Wu2025}.
Related experiments report acceptor brightening associated with polariton
relaxation in semiconductor heterostructures \cite{Hu2024PolaritonTransfer}
and fluorescence signatures of transfer between spatially separated
molecular layers coupled to optical cavity modes
\cite{Kertzscher2026PolaritonTransfer}.
A shared mode can also mediate Transfer without itself acquiring appreciable
real population.
In a dispersive photonic architecture, donor and acceptor can remain
resonant with one another while both are detuned from the cavity mode.
Virtual-photon exchange then generates an effective donor--acceptor
interaction without requiring the transferred excitation to be stored as
a real cavity photon \cite{Blais2007CircuitQED}. For fixed coupling between
each endpoint and the mode, increasing the detuning reduces mediator
participation and can suppress loss through that mode, but generally
weakens the effective interaction and slows transfer. Experimental examples
include spectroscopically established cavity-mediated exciton hopping
\cite{Husel2026Dispersive} and time-domain coherent excitation exchange
between semiconductor spin qubits \cite{Dijkema2025iSWAP}.

A third approach is to make multiple donor and acceptor sites
participate coherently in transfer. Their transfer amplitudes can
reinforce or cancel one another \cite{Cao2022GeneralizedRET}.
In ideal phase-matched models with comparable pair couplings, even a single excitation 
coherently delocalized over $N_D$ donor sites and $N_A$ acceptor
sites can enhance the transfer rate by up to $N_DN_A$ relative to
a single pair, an effect known as supertransfer
\cite{Kushwaha2025Supertransfer}. Sharma \emph{et al.} grew highly ordered pentacene monolayers and observed rapid exciton transport at room temperature,
which they attributed to supertransfer arising from coherent
delocalization \cite{Sharma2020Supertransport}.

Finally, Transfer can be modified by applying external fields
or engineering the environment.
Allcock \emph{et al.} predict that an auxiliary off-resonant laser
can change the rate of donor--acceptor energy transfer without
supplying the excitation energy being transferred
\cite{Allcock2000LARET}. In coupled networks, destructive interference can also trap excitation
in states that do not connect to the intended output. Suitable
perturbations of site energies or local dephasing can lift this blocking
and improve delivery---a process referred to as noise-assisted transport \cite{Caruso2009NoiseAssisted}. Noise-assisted
transport has been demonstrated experimentally in both optical cavity
networks and superconducting circuits
\cite{Viciani2015NoiseAssisted,Potocnik2018LightHarvesting}.

At nuclear scales, collective excitation and coherent control are already
established in nuclear quantum optics. Haber \emph{et al.} demonstrated
coherent Rabi dynamics involving spatially separated nuclear ensembles
\cite{Haber2017}.
More recently, collective resonant excitation and emission into a shared
single-mode hard-x-ray waveguide have been demonstrated for M\"ossbauer
nuclei \cite{Lohse2025}, and theory explores collective nuclear polaritons
with enhanced and tunable light--nucleus coupling \cite{Yang2026}.

\subsection{Transform: from altered dynamics to altered outcomes}
\label{sec:transform}

\textit{Transform} describes operations in which energy comes to reside in a
different physical degree of freedom or configuration, for example:
\begin{equation}
    A^* \xrightarrow{\mathrm{Transform}} C,
    \qquad
    A^* + B \xrightarrow{\mathrm{Transform}} C + D .
\end{equation}
This includes dissociation of a bound excitation into separated charge,
driven structural changes, and formation of chemical or nuclear reaction
products. Excite or Transfer can supply the initial population of excited states; 
Transform describes the subsequent change in the carrier or physical
configuration. Ordinary chemical reactions and charge separation also
perform this function.

Two forms of Transform are especially useful to distinguish: changing the energetic carrier and steering which reaction pathway or product channel receives the energy. 
Quantum engineering enters by modifying the states, couplings or environments that govern these transformations.

Changing the energetic carrier is central when energy initially stored in one degree of freedom must come to reside in another.
In organic light-harvesting and photovoltaic materials, optical excitation
commonly produces a bound electron--hole pair (excitons) but this is only useful it we can convert the exciton to charge. Ivanovi\'c
\emph{et al.} found in simulations that stronger intermolecular electronic
coupling or dielectric screening opens dissociation pathways through
transiently delocalized states with mixed exciton and charge-transfer
character \cite{Ivanovic2025}. Hymas \emph{et al.} incorporated
donor--acceptor and charge-transport layers into an organic microcavity,
connecting optical excitation to charge separation and measured electrical
output \cite{Hymas2026}.
At nuclear energies, Kekez \emph{et al.} reported evidence for a
decay pathway in $^{93\mathrm{m}}$Nb that divides the nuclear
excitation energy between excitation of a bound electron and
photon emission \cite{Kekez1985ElectronicBridge}.

A second form of Transform is to steer which transformation occurs when several
reaction pathways or products are available.
Beyond conventional chemical and thermodynamic control, confined electromagnetic modes provide another means of modifying molecular states and energy-flow pathways. Brawley \emph{et al.} observed dehydration of the reacting material at a lower heating-stage temperature on resonant nanocavity arrays, attributing the effect to enhanced radiative energy transfer \cite{Brawley2025}.
Experiments have reported changes in chemical product proportions
when molecular vibrations are strongly coupled to optical cavities
\cite{Thomas2019Selectivity,Sau2021Stereoselectivity}.
Models explore how cavity tuning can alter reaction rates and
product branching \cite{YingHuo2024,Mondal2026}.
Such effects depend on the reaction and coupling conditions;
some experiments find no substantial rate change \cite{Muller2025}.

\begin{figure*}[!t]
    \centering
    \includegraphics[width=0.76\linewidth]{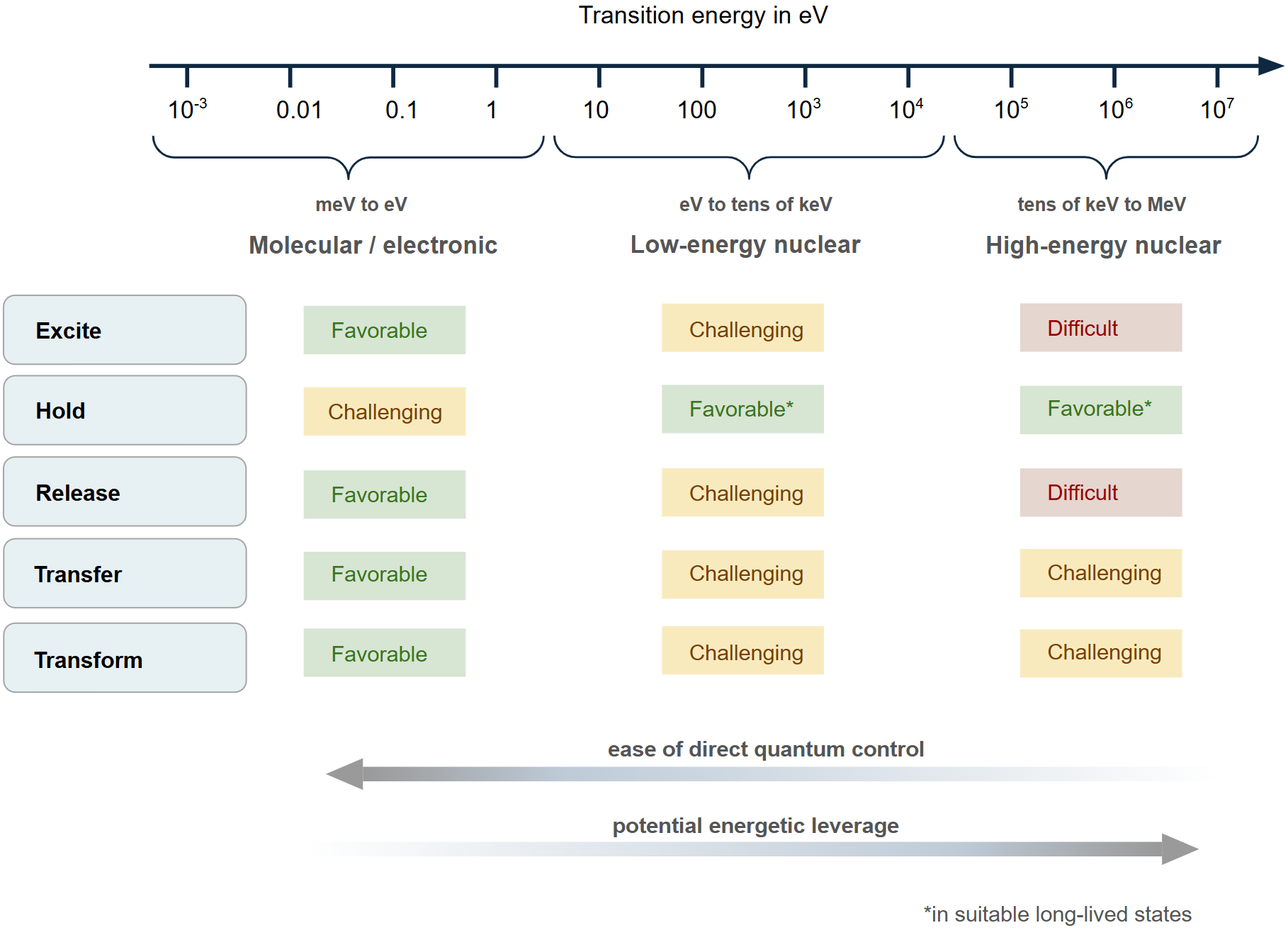}
    \caption{Implementation difficulty across transition-energy regimes. Energy ranges are schematic. Colors indicate qualitative difficulty for representative systems. The lower arrows indicate broad trends toward more
demanding direct control and greater potential energetic leverage at
higher transition energies.}
    \label{fig:scales}
\end{figure*}

At nuclear scales, the analogous question is whether preparation of the
reacting state or its environment can alter the distribution among competing
reaction channels.
Deuteron--deuteron (d--d) fusion provides a simple
example through the strong-interaction channels \(d(d,p)t\) and
\(d(d,n){}^3\mathrm{He}\). Nuclear spin preparation provides one possible handle on reaction outcomes. Measurements with polarized deuterons show spin-dependent changes in the angular distributions of the d--d fusion products, while calculations predict that spin alignment can modify the rates of the proton- and neutron-producing channels by different amounts, hence altering their branching ratio \cite{Paetz1992PolarizedDD,Deltuva2010PolarizedDD,Viviani2023DD}. Huke \emph{et al.} reported a host-material dependence of the
neutron-to-proton branching ratio in low-energy d--d fusion.
Their model reproduced the trend by assuming partial spin
polarization, but the study did not establish how the host would
generate that polarization \cite{Huke2006}.
More recently, Dubey \emph{et al.} reported signatures of an additional
electron--positron-pair emission channel in very-low-energy d--d fusion, with its
branching relative to the proton and neutron channels increasing toward
lower projectile energy \cite{Dubey2025}. 
These studies motivate investigation of nuclear state
preparation, the exploitation of nuclear resonances and the material environment as variables
that may alter the distribution of reaction outcomes.

\subsection{The same functions across energy scales}

The same five functions apply across physical platforms, but they are not
equally easy to implement. Figure~\ref{fig:scales} compares this difficulty
across transition energy regimes. Molecular and electronic excitations in the meV--eV range are generally the most directly addressable: Excite, Release, Transfer and Transform can often be implemented with optical, microwave, electrical or material couplings, while long retention often requires additional protection.

Nuclear systems span a much broader range. Exceptional low-energy transitions such as the approximately 8.4~eV \(^{229}\)Th isomer \cite{Zhang2024,Hiraki2024}, together with Mössbauer transitions such as the 14.4~keV transition in \(^{57}\)Fe and the \(12.4~\mathrm{keV}\) transition in \(^{45}\)Sc, occupy an intermediate regime in which direct excitation and controlled release are more demanding 
but remain experimentally accessible. Long-lived nuclear states can make Hold particularly favorable once population has been prepared or is already available, as illustrated by the exceptionally metastable \(^{180m}\)Ta state at approximately 77 keV. At higher nuclear transition energies, direct excitation and especially controlled Release generally become more difficult because suitable resonant fields, spectral brightness and coupling strengths are harder to obtain, while metastable states can still provide excellent intrinsic retention. In that regime, Transfer and Transform through shared modes, collective states and suitable intermediate states offer prospective alternatives to directly addressing the nuclear transition with a resonant field.

Figure~\ref{fig:scales} suggests a design strategy: exploit the functions that a platform provides naturally and engineer the missing ones.

\section{From functions to capabilities and applications}

We now assess the capabilities Storage, Routing and Conversion through the performance
required by corresponding applications. Table~\ref{tab:capabilities} summarizes their functional basis and characteristic performance metrics. These three examples
illustrate how functions can be combined into capabilities across
different physical platforms.

Each capability is described by multiple performance metrics whose relative
importance depends on the application. Long retention can come at the expense of rapid release; high routing selectivity
can come at the expense of throughput; and high conversion yield can require
slower operation or greater control energy. The following subsections
examine these trade-offs through representative implementations and application
benchmarks.

\begin{table}[!h]
\centering
\caption{Representative functional combinations and characteristic performance metrics for Storage, Routing and Conversion.}
\label{tab:capabilities}

\begingroup
\footnotesize
\linespread{1.10}\selectfont

\begin{tabularx}{\linewidth}{
    >{\bfseries\raggedright\arraybackslash}p{0.12\linewidth}
    >{\raggedright\arraybackslash}p{0.32\linewidth}
    >{\raggedright\arraybackslash}X
}
\toprule
\textbf{Capability} &
\textbf{Representative functional basis} &
\textbf{Characteristic performance metrics} \\
\midrule

Storage &
Excite $\rightarrow$ Hold $\rightarrow$ Release,
sometimes with Transfer &
\begin{tabitemize}
\item Usable energy and energy density
\item Retention
\item Charging/release power
\item Recovery and round-trip efficiencies
\item Durability
\end{tabitemize}
\\
\addlinespace[1.2em]

Routing &
Transfer, often with Excite and/or Release &
\begin{tabitemize}
\item Routing efficiency $\eta_{\mathrm{route}}$
\item Destination selectivity $S_{\mathrm{route}}$
\item Control leverage $L_E$ or $L_P$
\item Loss and speed
\item Access to otherwise weak or uncontrollable pathways
\end{tabitemize}
\\
\addlinespace[1.2em]

Conversion &
Transform, often following Excite or Transfer &
\begin{tabitemize}
\item Conversion efficiency $\eta_{\mathrm{conv}}$
\item Product selectivity $S_P$
\item Throughput
\item Control overhead
\item Unwanted-channel suppression
\item Durability
\end{tabitemize}
\\

\bottomrule
\end{tabularx}

\endgroup
\end{table}

\subsection{Storage: retaining energy and making it available when needed}

A common implementation of the Storage capability combines Excite, Hold and Release. Quantum battery
research has usually framed this sequence in terms of charging, capacity, and
extractable work; recent reviews also emphasize open system behavior,
extraction, implementation and scaling \cite{Campaioli2024RMP,Ferraro2026NRP}.
The following paragraphs give an overview of key performance requirements.

\paragraph{Usable energy and energy density.}
Conventional batteries provide a reference scale. High-energy
lithium-ion cells have reached cell-level specific energies of roughly $300~\mathrm{Wh\,kg^{-1}}$ \cite{Wang2022FastCharging}.
Automotive applications assess this alongside charging and discharging
performance, round-trip efficiency, cycle life and durability
\cite{USABC2025}. Molecular-microcavity quantum batteries occupy a
different region of the design space. In the experiment of Quach
\emph{et al.}, the largest strongly coupled sample contained about
$1.6\times10^{11}$ Lumogen-F orange (LFO) molecules, with an estimated peak stored energy of
$0.108~\mathrm{eV}$ per molecule obtained from fits to the measured dynamics
\cite{Quach2022}. Normalizing this energy to the active LFO molecules gives
about $4.1~\mathrm{Wh\,kg^{-1}}$. Including the polymer host, with LFO
constituting 10\% of the film mass, gives about
$0.41~\mathrm{Wh\,kg^{-1}}$. These values describe stored excitation;
recovery into a useful output was not measured in that experiment. The
film-level specific energy is orders of magnitude below the lithium-ion
cell benchmark, and including the mirrors, substrate and other inactive
components would lower the device-level value further. The significance of the experiment lies not in its absolute stored energy, but in the observed collective enhancement of energy absorption, i.e., in the charging performance.

\paragraph{Charging and discharging performance.}
Collective coupling provides a route to very rapid charging and release.
The mentioned Quach \emph{et al.} experiment demonstrated sub-picosecond
charging dynamics with a charging power per molecule that increased with
ensemble size \cite{Quach2022}. A distinct regime arises when energy is
accumulated comparatively slowly and released rapidly. Wang \emph{et al.} propose dynamically controlling the output coupling of a spin--cavity system to release stored energy in nanosecond, watt-level coherent microwave bursts \cite{Wang2026}. For comparison, even LiFePO$_4$ electrodes optimized for fast discharge operate on a full-discharge-equivalent timescale of order $10~\mathrm{s}$ \cite{Kang2009Ultrafast}.
Waveguide-QED proposals combine subradiant storage with controlled
superradiant release \cite{Chen2026}. For a charge--store--discharge cycle,
temporal concentration can be characterized by
$C_P=P_{\mathrm{peak,out}}/\overline{P}_{\mathrm{charge}}$, where
$\overline{P}_{\mathrm{charge}}$ is the mean input power over the charging
interval at a stated system boundary. Pulse energy, duration and the
recovered-energy fraction should accompany this ratio.

\paragraph{Retention times.}
Stored energy must remain accessible for the timescale required by the
application. Retention may be characterized by a lifetime
$\tau_{\mathrm{ret}}$, a self-discharge rate or a retention curve. Tibben
\emph{et al.} transferred excitation from a collectively coupled absorbing
layer into molecular triplet states and reported a lifetime of
$40.3\pm0.4~\mu\mathrm{s}$, roughly three orders of magnitude longer than
previous molecular-microcavity quantum-battery demonstrations
\cite{Tibben2025}. Hymas \emph{et al.} similarly use metastable molecular
states whose populations persist long after ultrafast charging
\cite{Hymas2026}. Despite this progress, the retention times of present molecular quantum batteries remain many orders of magnitude shorter than those of conventional energy-storage technologies designed for long-duration storage, such as electrochemical batteries.

\paragraph{Nuclear storage.}
Nuclear storage is appealing as it promises both much higher energy density than conventional storage and very long retention times, when long-lived nuclear states are used. The \(77~\mathrm{keV}\) excitation of \(^{180\mathrm{m}}\)Ta gives a useful sense of scale. At its natural abundance of about \(0.012\%\), the stored isomer energy corresponds to roughly \(1.4~\mathrm{kWh\,kg^{-1}}\) of natural tantalum, while pure \(^{180\mathrm{m}}\)Ta would contain about \(11.5~\mathrm{MWh\,kg^{-1}}\) in the isomer excitation alone. Even a \(5.45\%\)-enriched sample, an enrichment that has been used experimentally \cite{Belic2002Ta180m}, would correspond to roughly \(0.63~\mathrm{MWh\,kg^{-1}}\).
The difficulty is gaining controlled access to this stored energy. For naturally occurring isomers such as \(^{180\mathrm{m}}\)Ta, the stored excitation is already present; the central challenge is therefore controlled Release, as discussed above. A rechargeable implementation would additionally require a practical way to achieve Excite. The combination of very high stored energy and very long retention therefore motivates efforts to gain greater control over transitions of long-lived nuclear states.

\newpage

\subsection{Routing: controlling where energy goes}

Routing is centered on controlled Transfer. The destination can be another
molecule, circuit element or nuclear species, or an internal degree of freedom
such as a vibrational mode, spin state or reaction coordinate. A useful router must do more than demonstrate the physical function of Transfer: it must perform well on the metrics that matter for the intended application, which can include efficiency, destination selectivity, transfer rate, and control cost.

\paragraph{Routing efficiency.}
A first performance metric is routing efficiency,
\(\eta_{\mathrm{route}}=E_{\mathrm{target}}/E_{\mathrm{route,in}}\),
with the input and target energies defined for the chosen system boundary.
Molecularly designed light-harvesting architectures provide benchmarks. Kodis \emph{et al.} reported an essentially unity quantum yield
for antenna-to-porphyrin excitation transfer, followed by a
charge-separation yield of 96\% in the zinc-containing complex
\cite{Kodis2006}. In a patterned biohybrid platform, Rousseau \emph{et al.}
reported exciton diffusion over about \(200~\mathrm{nm}\) and approximately
30\% inter-complex transfer efficiency within overlapping antenna and
reaction-center-containing regions \cite{Rousseau2025}. These reported
yields describe how often excitation transfer or charge separation succeeds.
To compare them with an energy-based routing efficiency, one must also know
how much energy is retained in each successful event.

\paragraph{Energetic leverage.}
As noted under Transfer, the control need not supply the energy being routed. We define the energetic leverage
\(L_E=E_{\mathrm{transfer}}/E_{\mathrm{control}}\), with an analogous power
ratio for continuous operation. Large leverage is familiar from relays and
transistors. For microscopic mode-mediated transfer, the separation between the energy of the transferred
excitation and the characteristic quantum energy of its mediator is particularly important. We
characterize this separation by
\(\chi_E=E_{\mathrm{exc}}/(\hbar\omega_{\mathrm{med}})\), where
\(E_{\mathrm{exc}}\) is the energy of an individual excitation and
\(\omega_{\mathrm{med}}\) is a specified characteristic frequency of the
mediating mode or spectrum. The two ratios address different aspects: energetic leverage compares the
energy transferred with the expenditure needed to control it; energy-scale
separation compares an individual excitation quantum with the characteristic
quantum energy of the mediator. As excitation energies increase, using a mediating bus near the excitation frequency becomes progressively harder. Hard-x-ray cavities and waveguides have been realized at energies of order tens of keV \cite{Haber2017,Lohse2025}, but in the hundreds-of-keV to MeV regime practical resonant cavities with useful confinement, coupling efficiency and throughput are essentially unavailable. Transfer with \(\chi_E \gg 1\) could therefore open a qualitatively different design space, in which a much lower-energy mode mediates the interaction without requiring a resonator near the energy of the excitation itself.
Early theoretical work provides precedents for this energy-scale
separation. Hagelstein and Chaudhary considered excitation transfer
between two two-level systems coupled through a common oscillator whose
quantum energy was much smaller than the transition energies \cite{Hagelstein2008ExcitationTransfer}.
Nagibarova and Nagibarov considered virtual-phonon-mediated electronic
transfer with an illustrative optical excitation energy approximately
fifty times the Debye-cutoff phonon energy \cite{NagibarovaNagibarov1969}.
DeLosh and Grant treated transfer between impurity ions in the regime
\(E_{\mathrm{exc}}\gg\hbar\omega_{\mathrm{ph,max}}\)
\cite{DeLosh1970}. In these models, the low-energy mediator generates the transfer interaction while the donor supplies the excitation energy. These results motivate the
search for transfer architectures that combine substantial energy-scale separation with useful transfer rates. 

\paragraph{Selectivity.}
When several destinations compete for excitation, routing selectivity
characterizes the fraction of delivered energy reaching the intended
target. A corresponding metric is
\(S_{\mathrm{route}}=E_{\mathrm{target}}/\sum_j E_j\), where the sum
includes the target and the other specified output channels.
Reconfigurable routing allows the preferred
destination to be changed within the same device.
Experimental demonstrations span several forms of destination control.
Marsault \emph{et al.} used an optical control beam to shift a confined
polariton state and direct injected polariton pulses into either of two
output wires
\cite{Marsault2015PolaritonRouter}. Liu \emph{et al.} demonstrated
electrically programmable routing of interlayer excitons in bilayer
WSe$_2$, using gate voltages to direct propagation between selected
ports \cite{Liu2020ExcitonRouter}. In superconducting circuits, Miao
\emph{et al.} implemented a router in which the state of a switch qubit
selects which of two output qubits receives the input state, including
coherent routing controlled by a superposed switch state \cite{Miao2025QuantumRouter}.
The implementations differ, but the common task is destination selection
using optical, electrical or quantum state controls. 
To compare
these demonstrations using \(S_{\mathrm{route}}\), the measured observables
must be related to the energy delivered at each output.

\paragraph{Access to new pathways.}
Some applications are limited by the absence of a useful connection between source and
destination. Quantum engineering can help create such connections.
Spatially separated systems provide one example. Conventional exchange
operations between semiconductor spin qubits rely on wavefunction overlap
between neighboring quantum dots, typically separated by
\(100\)--\(200~\mathrm{nm}\). Dijkema \emph{et al.} instead connected spin qubits \(250~\mu\mathrm{m}\)
apart through a shared resonator, enabling population exchange in
approximately \(42~\mathrm{ns}\) at separations far beyond the usual
overlap-based interaction \cite{Dijkema2025iSWAP}.
Pathway access can also arise through changes in the available excited
states. In the dye pair studied by Dovzhenko \emph{et al.}, excitation
normally flows from the higher-energy donor to the acceptor. Cavity
coupling created a mixed donor--light state below the acceptor in energy,
allowing excitation to flow from the acceptor into this new state
\cite{Dovzhenko2021RoleReversal}.
Polak \emph{et al.} observed long-lived polariton emission attributed to
an additional pathway from normally dark triplet-pair populations
\cite{Polak2020TripletHarvesting}. Here, the additional pathway gives excitation stored in normally
non-emissive triplet-pair states access to a light-emitting output through
mixing with the cavity-coupled singlet states.

\paragraph{The nuclear frontier.}
Nuclear systems provide an extreme case of this Routing design space. If a mode with a characteristic quantum energy of \(1~\mathrm{eV}\)
mediated transfer of a \(14.4~\mathrm{keV}\) nuclear excitation, the
energy-scale separation would be about four orders of magnitude,
corresponding to \(\chi_E \sim 10^4\), with still larger scale separations for
lower-energy mediators or higher-energy nuclear transitions.
A low-energy electromagnetic, electronic, vibrational or magnetic degree of
freedom could, in principle, govern a spatially extended pathway through which
pre-existing nuclear-scale energy becomes available at a selected acceptor.
Routing can go beyond an excitation swap between otherwise unchanged nuclei:
energy transfer can also accompany a change in nuclear configuration or identity,
as in neutron-transfer reactions
\cite{Timofeyuk2020Transfer,Jones2022NeutronTransfer}.
Collective participation could strengthen a weak transfer interaction
where the required state structure and coupling geometry permit it. 
Collective nuclear excitation and coherent control have been demonstrated
using near-resonant x-ray fields and engineered photonic structures
\cite{Haber2017,Heeg2021,Lohse2025}. A complementary experimental
starting point is local electronic--nuclear energy exchange: Kishimoto
\emph{et al.} observed nuclear excitation in \(^{197}\)Au following
x-ray preparation of atomic vacancies and subsequent electronic
relaxation \cite{Kishimoto2000NEET}.
On the theory side, Tang \emph{et al.} predict shallow neutron bound states extending over tens of nanometers, with wavefunctions shaped by nanocrystal geometry and
nuclear-spin polarization \cite{Tang2024NeutronStates}. A neutron-based
router would additionally require a mechanism connecting these shallow
states to a specified nuclear source and receiver. 
Other theoretical work considers transfer between nuclear configurations
through shared phononic, plasmonic and magnetic degrees of freedom,
including possible collective enhancement
\cite{Hagelstein2025Models}. Experimental validation would require showing that controlled changes in the
proposed coupling produce energy delivery at a specified
receiver. An ambitious nuclear routing objective is to establish a
controllable source--receiver pathway that delivers nuclear energy to
a selected output at a useful rate and with high energetic leverage.

\subsection{Conversion: controlling what useful outcome results}
\label{sec:conversion}

Conversion is centered on Transform, usually after Excite or Transfer has
delivered energy to the relevant degree of freedom. Its performance is
assessed through the desired output: electrical energy, a chemical or
nuclear reaction product, or a functional material state. Efficiency,
selectivity, yield and throughput must therefore be considered alongside
the control expenditure and operating conditions required to achieve them.

\paragraph{Energy/carrier conversion.}
When input and useful output are measured as energy, a conventional
conversion efficiency is appropriate,
\(\eta_{\mathrm{conv}}=E_{\mathrm{useful,out}}/E_{\mathrm{input}}\),
with both energies evaluated over a stated system boundary.
Photovoltaics provide a mature benchmark for converting absorbed optical excitation into separated charge and electrical output. State-of-the-art photovoltaic cells achieve efficiencies of tens of percent \cite{Green2026EfficiencyTables}, including organic cells above 20\% \cite{Wang2024NFA}. 
In an organic microcavity, Hymas \emph{et al.} reported roughly
threefold higher external quantum efficiency than corresponding
no-cavity devices. Current--voltage measurements also showed
an electrical-power enhancement that increased with the number
of absorbing molecules under steady illumination \cite{Hymas2026}.
External quantum efficiency counts collected electrons per photon,
not the fraction of optical energy converted to electricity.

\paragraph{Yield, selectivity and branching.}
For chemical or nuclear transformations, a measure of product
selectivity is \(S_P=N_{\mathrm{desired}}/\sum_k N_k\), where \(N_k\)
counts events in channel \(k\).
Absolute yield and throughput are reported
separately. For two competing channels, the ratio \(B_{i/j}=N_i/N_j\)
compares their relative contributions, expressed below as a
chemical product ratio or a nuclear branching ratio.
Sau \emph{et al.} reported a change in the product ratio of a
thermal ring-opening reaction from approximately $13{:}1$ to
$28{:}1$ when an optical cavity was tuned to a selected molecular
vibration. The accompanying decrease in reaction rate illustrates
a selectivity--rate trade-off \cite{Sau2021Stereoselectivity}. 
The nuclear reaction studies introduced in the Transform subsection likewise motivate nuclear state preparation and the material environment
as candidate controls for nuclear branching ratios \cite{Paetz1992PolarizedDD,Huke2006,Dubey2025}.
A relevant Conversion capability would require a reproducible shift in
yields.

\begin{table*}[!h]

\centering

\caption{Illustrative stretch scenarios for quantum energy applications over a 10--15-year horizon. Targets describe proposed performance in a chosen implementation, while the cited research 
provides individual experimental ingredients or theoretical approaches. 
The final column includes longer- term possibilities requiring advances 
beyond the listed targets and potentially beyond this horizon.}

\label{tab:success}

\begingroup
\footnotesize
\linespread{1.10}\selectfont
\setlength{\tabcolsep}{4pt}
\renewcommand{\arraystretch}{1.10}
\setlength{\emergencystretch}{1em}

\begin{tabularx}{\linewidth}{@{}
    >{\raggedright\arraybackslash}p{2.9cm}
    >{\raggedright\arraybackslash}X
    >{\raggedright\arraybackslash}X
    >{\raggedright\arraybackslash}X@{}}
\toprule
\textbf{Application / capability} &
\textbf{Literature anchor} &
\textbf{Stretch performance targets} &
\textbf{Potentially enabled technology} \\
\midrule

\textbf{Low-heat quantum processor drives}\par
\textit{Storage}
& Shared quantum energy stores for repeated qubit driving with fewer microwave drive lines (theory) \cite{Kurman2026}
& Local energy reuse across \(\mathord{>}10\) qubits; \(10\times\) lower average drive-related cryogenic heat and removal of most dedicated microwave drive lines.
& Larger quantum computing processors within a fixed cooling and wiring budget \\
\addlinespace[8pt]

\textbf{Nuclear quantum batteries}\par
\textit{Storage}
& Nuclear excitation and lifetime control (experiment)
  \cite{Shvydko2023Scandium,Schaden2025};
  collective charging and release (theory)
  \cite{Dahal2026,Yang2026}
& Recoverable energy per gram of complete device \(\mathord{>}100\times\) above electrochemical batteries after \(\mathord{\gtrsim}1\) s of storage—or years in shelf-life designs—with on-demand recovery of \(\mathord{\gtrsim}50\%\) of the stored energy.
& High-density energy storage with output on demand; long-shelf-life
  sources for mass-constrained and infrequently serviced systems \\
\addlinespace[8pt]

\textbf{Artificial photosynthesis}\par
\textit{Routing}
& Efficient antenna delivery, integrated reaction-center networks
and reconfigurable molecular pathways (experiment)
\cite{Wu2025,Kodis2006,Rousseau2025,Straight2005DoubleThrow}
& Reconfigurable delivery of absorbed energy across multistage molecular networks to chosen reaction centers or catalytic sites, with energy delivery efficiency \(\mathord{>}95\%\) and destination selectivity \(\mathord{>}99\%\).
& Adaptable solar fuel and chemical production systems that redirect
absorbed energy to reaction sites, integrating light harvesting and chemical processing for increased efficiency \\
\addlinespace[8pt]

\textbf{Nuclear energy routing}\par
\textit{Routing}
& Collective and electronic--nuclear interfaces (experiment)
  \cite{Haber2017,Kishimoto2000NEET};
  material-mediated transfer and extended neutron states (theory)
  \cite{Hagelstein2025Models,Tang2024NeutronStates}
& Switchable delivery of keV–MeV energy to selected receiver systems using eV-scale or lower-energy controls; delivered/control power ratio \(\mathord{>}1000\); ultimately watt-scale power at the receiver side.
& Targeted nuclear energy delivery within microscopic networks and transfer-assisted isomer depletion;
  longer-term potential for material-mediated nuclear energy extraction \\
\addlinespace[8pt]

\textbf{Selective chemistry}\par
\textit{Conversion}
& Reported cavity-dependent changes in product proportions (experiment)
  \cite{Thomas2019Selectivity,Sau2021Stereoselectivity};
  models of reaction-rate and branching control (theory)
  \cite{YingHuo2024,Mondal2026}
& Shift roughly 50:50 branching to \(>95\%\) desired product, with continuous production of high-value chemicals.
& Selective production of desired chemicals with fewer unwanted
  byproducts and less downstream separation \\
\addlinespace[8pt]

\textbf{Nuclear reaction channel control}\par
\textit{Conversion}
& Spin-dependent fusion observables and host- and energy-dependent
  product branching (experiment)
  \cite{Paetz1992PolarizedDD,Huke2006,Dubey2025};
  calculations and proposed acceptor-assisted pathway engineering
  (theory)
  \cite{Deltuva2010PolarizedDD,Viviani2023DD,Hagelstein2025Models}
& Reproducibly increase a desired-to-unwanted reaction channel ratio by \(\mathord{>}10\times\) without reducing the desired rate under
matched conditions.
& Product-selective nuclear reactions; potential fusion systems with
  reduced neutron output \\
\bottomrule
\end{tabularx}

\endgroup
\end{table*}

\section{What could success look like?}

The preceding sections identify energetic functions, their physical
implementations and the metrics that determine their usefulness. We now
consider what combinations of performance could make these capabilities
technologically consequential. Table~\ref{tab:success} presents six application-level stretch
scenarios aimed at a 10--15-year horizon, with two examples
each of Storage, Routing and Conversion.
These scenarios are conditional engineering objectives: their realization requires further progress in integration, scaling and
operating cost, and in some cases validation of a proposed microscopic
pathway. The cited studies provide individual
experimental results or theoretical proposals as anchoring points for projected progress.

\paragraph{Storage: local reuse and high-specific-energy resources.}
The two Storage scenarios illustrate different sources of value. For a
quantum computing processor, a small inventory of microwave excitations could be
reused through many operations. Kurman \emph{et al.} predict that replacing
individual microwave drives with shared precharged modes could increase the
qubit capacity of a cryogenic system by up to about $4\times$ under
favorable wiring and thermal assumptions \cite{Kurman2026}. The intended benefit is more useful computation within a fixed cooling and
interconnect budget. Assessing that benefit requires including preparation,
error protection and replenishment in the comparison.

Nuclear quantum batteries instead seek to combine large excitation energies
with controlled retention and retrieval. Possible implementations include
charging intrinsically longer-lived levels, protecting otherwise short-lived
excitations, and triggering discharge of a long-lived isomer inventory.
Resonant excitation of the $12.4$-keV state of $^{45}$Sc, with a mean lifetime
of approximately $0.47~\mathrm{s}$, provides an experimental starting point
for deliberately charged nuclear stores \cite{Shvydko2023Scandium};
collective charging and tunable release are being explored theoretically
\cite{Dahal2026,Yang2026}. Engineered Hold must address all dominant decay
channels: suppressing collective radiation alone leaves local losses such
as internal conversion \cite{Shvydko1996,Dahal2026}. 
If substantial energy could be recovered after the required retention time, nuclear stores could be valuable in mass-constrained or difficult-to-service systems even if their cost per kilowatt-hour remained higher than that of incumbent energy-storage technologies.

\paragraph{Routing: connecting resources to useful destinations.}
Artificial photosynthesis gives programmable Routing a concrete purpose:
collect excitation across an extended absorber network and direct it toward
reaction centers or catalytic sites able to use it. Efficient molecular
antennas, integrated protein networks and switchable molecular pathways
supply distinct ingredients \cite{Kodis2006,Rousseau2025,Straight2005DoubleThrow}.
Successful integration could couple light harvesting more directly to induced chemical reactions, reducing losses between absorption and reaction and improving the overall efficiency of converting incident light into desired chemical products.

Transfer-assisted isomer depletion could provide a controlled outlet for nuclear excitation that is difficult to release directly. A more far-reaching possibility is to couple an exothermic nuclear transformation to accepting states, enabling material-mediated release of nuclear binding energy \cite{Hagelstein2025Models}. If such a pathway supported sustained net output, it could open a different route to nuclear energy extraction and also provide a means of influencing the final products by routing energy to acceptor states with favorable subsequent decay channels.

\paragraph{Conversion: selecting outcomes and changing system costs.}
Chemical product control could create value by reducing the need for
purification, recycling and waste management. Moving from $50{:}50$ to
$95{:}5$ desired/unwanted branching, for example, reduces unwanted product per unit of
desired product by a factor of nineteen. For manufacturing, the challenge is to retain such a selectivity advantage while achieving significant yield and throughput at an acceptable cost. Cavity-modified
rate and branching theories identify candidate interventions
\cite{YingHuo2024,Mondal2026}; their technological value depends on the performance of the complete process relative to competing synthetic methods.

Nuclear reaction channel control addresses a different system constraint:
the products and energy carriers generated by a transformation. In a
compatible reaction and fuel cycle, an order-of-magnitude reduction in
damaging neutron exposure per unit of fusion output could ease
component replacement and maintenance. Neutron fluence limits, component replacement costs
and plant uptime enter directly into fusion economic viability
\cite{Whyte2026Economics}. Realizing that benefit would require maintaining
useful reaction power while including the cost of preparing
and controlling the reacting states. The larger opportunity is to change
which reactor designs become economically attractive.

These six scenarios are illustrative. Further opportunities include chip-scale
coherent microwave sources for pulsed spectroscopy and local spin control
\cite{Kersten2023Triggered,Wang2026}, and selectively driven material states
for reconfigurable optical, electronic or magnetic
functionality \cite{Huang2026}.

\section{Outlook: toward programmable quantum energy}

Established energy technologies are controlled mainly through macroscopic variables such as temperature, voltage and illumination.
Quantum energy engineering seeks to add control over the microscopic states, couplings and pathways that determine how energy is stored, transferred and transformed.

The next phase of quantum energy science is to test whether microscopic
control can improve complete devices under application-relevant conditions.
For quantum energy technologies to matter, they will ultimately need to
outperform conventional alternatives on application-relevant metrics,
including the full costs of preparation and control. An advantage can arise
from a combination of properties rather than from superiority on every
individual metric.

Experiments should test whether microscopic control remains useful under
realistic losses, operating conditions and control costs. Theory should
identify which favorable scalings survive disorder, competing loss channels
and finite control resources. Establishing where an effect ceases to be useful
helps identify viable architectures and shows which lessons can be transferred
to other platforms.

The examples discussed in this article also reveal shared research problems.
Preserving an energetic population, opening a controllable output and
maintaining useful transfer in the presence of loss recur in molecular,
photonic, superconducting and nuclear systems. Organizing research around
such recurring themes can help communities recognize where they face the same
engineering problem and exchange relevant design principles, without
assuming that the material-specific physics or attainable performance is
the same.

We use \emph{programmable quantum energy} to describe this systems-level
vision: deliberate microscopic control over how energy enters a system, how long it is
retained, where it is delivered, how it is released and which physical
changes it produces.

\section*{Acknowledgments}

The authors thank James Quach, Francesco Campaioli, Oliver Carefull, Geoff Enders, Erik Ziehm, Carlotta Barone-MacDonald, Eman Elshaikh, Colin Clark, Yet-Ming Chiang, Matt Trevithick, Joseph Peeraer, Richard Lester, Konrad Czerski, Ed Cruz, Bill Horak, Huw Price, Carl Page, and Frank Ling for helpful discussions.

The authors are grateful to ARPA-E and the Anthropocene Institute for financial support.

\section*{Author Contributions}

F.M. conceived and initiated the project and wrote the initial draft. M.K.L. edited and refined the manuscript in close coordination with F.M. All other authors reviewed the manuscript, provided comments and suggestions, and approved the final version.

\section*{Disclosure statement}

The authors report no competing interests. 

\newpage

\ifmethodsformat

    \bibliographystyle{metzler-preprint}

\else

    \clearpage
    \bibliographystyle{quantumenergy_initials}

\fi

\bibliography{references}

@article{Metzler2023,
  author  = {Metzler, Florian and Sandoval, Jorge I. and Galvanetto, Nicola},
  title   = {The emergence of quantum energy science},
  journal = {Journal of Physics: Energy},
  volume  = {5},
  pages   = {041001},
  year    = {2023},
  doi     = {10.1088/2515-7655/acfbb8}
}

@article{DiVincenzo2000,
  author  = {DiVincenzo, David P.},
  title   = {The physical implementation of quantum computation},
  journal = {Fortschritte der Physik},
  volume  = {48},
  pages   = {771--783},
  year    = {2000},
  eprint  = {quant-ph/0002077},
  archivePrefix = {arXiv},
  doi     = {10.1002/1521-3978(200009)48:9/11<771::AID-PROP771>3.0.CO;2-E}
}

@article{Binder2015,
  author  = {Binder, Felix C. and Vinjanampathy, Sai and Modi, Kavan and Goold, John},
  title   = {Quantacell: powerful charging of quantum batteries},
  journal = {New Journal of Physics},
  volume  = {17},
  pages   = {075015},
  year    = {2015},
  doi     = {10.1088/1367-2630/17/7/075015}
}

@article{Gyhm2022,
  author  = {Gyhm, Ju-Yeon and {\v{S}}afr{\'a}nek, Dominik and Rosa, Dario},
  title   = {Quantum charging advantage cannot be extensive without global operations},
  journal = {Physical Review Letters},
  volume  = {128},
  pages   = {140501},
  year    = {2022},
  doi     = {10.1103/PhysRevLett.128.140501}
}

@article{Quach2022,
  author  = {Quach, James Q. and McGhee, Kirsty E. and Ganzer, Lukas and Rouse, Dominic M. and Lovett, Brendon W. and Gauger, Erik M. and Keeling, Jonathan and Cerullo, Giulio and Lidzey, David G. and Virgili, Tersilla},
  title   = {Superabsorption in an organic microcavity: Toward a quantum battery},
  journal = {Science Advances},
  volume  = {8},
  pages   = {eabk3160},
  year    = {2022},
  doi     = {10.1126/sciadv.abk3160}
}

@article{Hu2026,
  author  = {Hu, Chang-Kang and Liu, Chilong and Zhao, Jingchao and others},
  title   = {Quantum Charging Advantage in Superconducting Solid-State Batteries},
  journal = {Physical Review Letters},
  volume  = {136},
  pages   = {060401},
  year    = {2026},
  doi     = {10.1103/sp5l-c6m8}
}

@article{Mochizuki2026,
  author  = {Mochizuki, Tatsuto and Takahashi, Shota and Sakurai, Atsunori and Sugimoto, Toshiki},
  title   = {Tip-Enhanced Vibrational Ladder Climbing in Surface Molecular System},
  journal = {arXiv preprint arXiv:2608.01698},
  year    = {2026},
  eprint  = {2608.01698},
  archivePrefix = {arXiv},
  doi     = {10.48550/arXiv.2608.01698}
}

@article{Zhang2024,
  author  = {Zhang, Chuankun and Ooi, Tian and Higgins, Jacob S. and others},
  title   = {Frequency ratio of the {$^{229\mathrm{m}}$Th} nuclear isomeric transition and the {$^{87}$Sr} atomic clock},
  journal = {Nature},
  volume  = {633},
  pages   = {63--70},
  year    = {2024},
  doi     = {10.1038/s41586-024-07839-6}
}

@article{Hiraki2024,
  author  = {Hiraki, Takahiro and Okai, Koichi and Bartokos, Michael and others},
  title   = {Controlling {$^{229}$Th} isomeric state population in a {VUV} transparent crystal},
  journal = {Nature Communications},
  volume  = {15},
  pages   = {5536},
  year    = {2024},
  doi     = {10.1038/s41467-024-49631-0}
}

@article{Dahal2026,
  author  = {Dahal, Pravin Kumar and Hymas, Kieran and Muir, Jack and Quach, James Q.},
  title   = {Collective Enhancement of Nuclear Excitation for a Nuclear Quantum Battery},
  journal = {arXiv preprint arXiv:2607.15319},
  year    = {2026},
  eprint  = {2607.15319},
  archivePrefix = {arXiv},
  doi     = {10.48550/arXiv.2607.15319}
}

@article{Tibben2025,
  author  = {Tibben, Daniel J. and Della Gaspera, Enrico and van Embden, Joel and Reineck, Philipp and Quach, James Q. and Campaioli, Francesco and G{\'o}mez, Daniel E.},
  title   = {Extending the Self-Discharge Time of {Dicke} Quantum Batteries Using Molecular Triplets},
  journal = {PRX Energy},
  volume  = {4},
  pages   = {023012},
  year    = {2025},
  doi     = {10.1103/bhyh-53np}
}

@article{Chen2026,
  author  = {Chen, Wei and Lin, Kuan-Ting and Lin, Guin-Dar and Jen, Hsiang-Hua},
  title   = {Scaling law for optimal excitation storage and superradiant release in waveguide {QED} systems},
  journal = {arXiv preprint arXiv:2607.17320},
  year    = {2026},
  eprint  = {2607.17320},
  archivePrefix = {arXiv},
  doi     = {10.48550/arXiv.2607.17320}
}

@article{Hymas2026,
  author  = {Hymas, Kieran and Muir, Jack B. and Tibben, Daniel and others},
  title   = {Superextensive electrical power from a quantum battery},
  journal = {Light: Science \& Applications},
  volume  = {15},
  pages   = {168},
  year    = {2026},
  doi     = {10.1038/s41377-026-02240-6}
}

@article{Wang2026,
  author  = {Wang, Yuanjin and Wu, Hao and Oxborrow, Mark and Zhao, Qing},
  title   = {Watt-level coherent microwave emission from dissipation-engineered solid-state quantum batteries},
  journal = {Physical Review Applied},
  volume  = {26},
  pages   = {014009},
  year    = {2026},
  doi     = {10.1103/2dw2-z21v}
}

@article{Cenedese2026,
  author  = {Cenedese, Gabriele and Benenti, Giuliano and Ferraro, Dario and Genoni, Marco G.},
  title   = {Boosting work extraction in quantum batteries via continuous environment monitoring},
  journal = {Physical Review Letters},
  volume  = {137},
  pages   = {050406},
  year    = {2026},
  doi     = {10.1103/f591-hx6x}
}

@article{Wu2025,
  author  = {Wu, Fan and Nguyen-Phan, Tu C. and Cogdell, Richard and others},
  title   = {Efficient cavity-mediated energy transfer between photosynthetic light harvesting complexes from strong to weak coupling regime},
  journal = {Nature Communications},
  volume  = {16},
  pages   = {5300},
  year    = {2025},
  doi     = {10.1038/s41467-025-60616-5}
}

@article{Haber2017,
  author  = {Haber, J. and Kong, X. and Strohm, C. and others},
  title   = {Rabi oscillations of {X}-ray radiation between two nuclear ensembles},
  journal = {Nature Photonics},
  volume  = {11},
  pages   = {720--725},
  year    = {2017},
  doi     = {10.1038/s41566-017-0013-3}
}

@article{Heeg2021,
  author  = {Heeg, Kilian P. and others},
  title   = {Coherent {X}-ray--optical control of nuclear excitons},
  journal = {Nature},
  volume  = {590},
  pages   = {401--404},
  year    = {2021},
  doi     = {10.1038/s41586-021-03276-x}
}

@article{Yang2026,
  author  = {Yang, Liufeng and Wang, Jinling and Li, Huijun and Cao, Junhui and Kavokin, Alexey and Wu, Congjun},
  title   = {Collective Nuclear Polaritons with Coherent and Tunable Excitation Dynamics},
  journal = {arXiv preprint arXiv:2603.13948},
  year    = {2026},
  eprint  = {2603.13948},
  archivePrefix = {arXiv},
  doi     = {10.48550/arXiv.2603.13948}
}

@article{YingHuo2024,
  author  = {Ying, Wenxiang and Huo, Pengfei},
  title   = {Resonance theory of vibrational strong coupling enhanced polariton chemistry and the role of photonic mode lifetime},
  journal = {Communications Materials},
  volume  = {5},
  pages   = {110},
  year    = {2024},
  doi     = {10.1038/s43246-024-00551-y}
}

@article{Mondal2026,
  author  = {Mondal, Subhadip and Kumar, Atul and Keshavamurthy, Srihari},
  title   = {Vibrational strong coupling influences product selectivity in a model for post transition state bifurcation reactions},
  journal = {arXiv preprint arXiv:2603.09713},
  year    = {2026},
  eprint  = {2603.09713},
  archivePrefix = {arXiv},
  doi     = {10.48550/arXiv.2603.09713}
}

@article{Huang2026,
  author  = {Huang, Junhan and Ju, Yongkang and Wang, Xinbo and others},
  title   = {Coherent seeding and control of dynamical ferroelectricity by phonon anharmonicity},
  journal = {arXiv preprint arXiv:2606.22896},
  year    = {2026},
  eprint  = {2606.22896},
  archivePrefix = {arXiv},
  doi     = {10.48550/arXiv.2606.22896}
}

@article{Dicke1954,
  author  = {Dicke, Robert H.},
  title   = {Coherence in Spontaneous Radiation Processes},
  journal = {Physical Review},
  volume  = {93},
  pages   = {99--110},
  year    = {1954},
  doi     = {10.1103/PhysRev.93.99}
}

@article{LloydMohseni2010,
  author  = {Lloyd, Seth and Mohseni, Masoud},
  title   = {Symmetry-enhanced supertransfer of delocalized quantum states},
  journal = {New Journal of Physics},
  volume  = {12},
  pages   = {075020},
  year    = {2010},
  doi     = {10.1088/1367-2630/12/7/075020}
}

@article{Higgins2014,
  author  = {Higgins, K. D. B. and Benjamin, S. C. and Stace, T. M. and Milburn, G. J. and Lovett, B. W. and Gauger, E. M.},
  title   = {Superabsorption of light via quantum engineering},
  journal = {Nature Communications},
  volume  = {5},
  pages   = {4705},
  year    = {2014},
  doi     = {10.1038/ncomms5705}
}

@article{Nickerson2021,
  author  = {Nickerson, Brenden S. and Pimon, Martin and Bilous, Pavlo V. and
             Gugler, Johannes and Kazakov, Georgy A. and Sikorsky, Tomas and
             Beeks, Kjeld and Gr{\"u}neis, Andreas and Schumm, Thorsten and
             P{\'a}lffy, Adriana},
  title   = {Driven electronic bridge processes via defect states in {$^{229}$Th}-doped crystals},
  journal = {Physical Review A},
  volume  = {103},
  pages   = {053120},
  year    = {2021},
  doi     = {10.1103/PhysRevA.103.053120}
}

@article{Schaden2025,
  author  = {Schaden, F. and Riebner, T. and Morawetz, I. and
             Toscani De Col, L. and Kazakov, G. A. and Beeks, K. and
             Sikorsky, T. and Schumm, T. and Zhang, K. and Lal, V. and
             Zitzer, G. and Tiedau, J. and Okhapkin, M. V. and Peik, E.},
  title   = {Laser-induced quenching of the {$^{229}$Th} nuclear clock isomer in calcium fluoride},
  journal = {Physical Review Research},
  volume  = {7},
  pages   = {L022036},
  year    = {2025},
  doi     = {10.1103/PhysRevResearch.7.L022036}
}

@article{Wang2022FastCharging,
  author  = {Wang, Chao-Yang and Liu, Teng and Yang, Xiao-Guang and Ge, Shanhai
             and Stanley, Nathaniel V. and Rountree, Eric S. and Leng, Yongjun
             and McCarthy, Brian D.},
  title   = {Fast charging of energy-dense lithium-ion batteries},
  journal = {Nature},
  volume  = {611},
  number  = {7936},
  pages   = {485--490},
  year    = {2022},
  doi     = {10.1038/s41586-022-05281-0}
}

@techreport{USABC2025,
  author      = {{United States Advanced Battery Consortium}},
  title       = {{USABC} Goals for Advanced Batteries for Light Duty Electric Vehicles ({EVs})},
  institution = {United States Council for Automotive Research},
  year        = {2025},
  note        = {Appendix A: USABC EV Performance Targets, version October 6, 2025},
  url         = {https://uscar.org/wp-content/uploads/2025/12/USABC-EV-RFPI.pdf}
}

@article{Kurman2026,
  author  = {Kurman, Yaniv and Hymas, Kieran and Fedorov, Arkady
             and Munro, William J. and Quach, James Q.},
  title   = {Powering Quantum Computation with Quantum Batteries},
  journal = {Physical Review X},
  volume  = {16},
  number  = {1},
  pages   = {011016},
  year    = {2026},
  doi     = {10.1103/l39v-jwwz}
}

@article{Kodis2006,
  author  = {Kodis, Gerdenis and Terazono, Yuichi and Liddell, Paul A.
             and Andr{\'e}asson, Joakim and Garg, Vikas and Hambourger, Michael
             and Moore, Thomas A. and Moore, Ana L. and Gust, Devens},
  title   = {Energy and Photoinduced Electron Transfer in a Wheel-Shaped
             Artificial Photosynthetic Antenna--Reaction Center Complex},
  journal = {Journal of the American Chemical Society},
  volume  = {128},
  number  = {6},
  pages   = {1818--1827},
  year    = {2006},
  doi     = {10.1021/ja055903c}
}

@article{Rousseau2025,
  author  = {Rousseau, Adrien and Richardson, Katherine H. and Nandy, Atanu
             and Vasilev, Cvetelin and Hoffmann, Madeline P. and Hunter, C. Neil
             and Johnson, Matthew P. and Schlau-Cohen, Gabriela S.},
  title   = {Exciton-Diffusion Enhanced Energy Capture in an Integrated
             Nanoscale Platform},
  journal = {ACS Nano},
  volume  = {19},
  number  = {15},
  pages   = {14865--14872},
  year    = {2025},
  doi     = {10.1021/acsnano.4c18713}
}

@article{Hagelstein2025Models,
  author  = {Hagelstein, Peter L. and Metzler, Florian and Lilley, Matt K.
             and Messinger, Jonah F. and Galvanetto, Nicola},
  title   = {Models for nuclear fusion in the solid state},
  journal = {arXiv preprint arXiv:2501.08338},
  year    = {2025},
  eprint  = {2501.08338},
  archivePrefix = {arXiv},
  doi     = {10.48550/arXiv.2501.08338}
}

@article{Campaioli2024RMP,
  author  = {Campaioli, Francesco and Gherardini, Stefano and Quach, James Q. and Polini, Marco and Andolina, Gian Marcello},
  title   = {Colloquium: Quantum batteries},
  journal = {Reviews of Modern Physics},
  volume  = {96},
  number  = {3},
  pages   = {031001},
  year    = {2024},
  doi     = {10.1103/RevModPhys.96.031001}
}

@article{Ferraro2026NRP,
  author  = {Ferraro, Dario and Cavaliere, Fabio and Genoni, Marco G. and Benenti, Giuliano and Sassetti, Maura},
  title   = {Opportunities and challenges of quantum batteries},
  journal = {Nature Reviews Physics},
  volume  = {8},
  pages   = {115--127},
  year    = {2026},
  doi     = {10.1038/s42254-025-00906-5}
}

@article{Shastri2025,
  author  = {Shastri, Rahul and Jiang, Chao and Xu, Guo-Hua and Venkatesh, B. Prasanna and Watanabe, Gentaro},
  title   = {Dephasing enabled fast charging of quantum batteries},
  journal = {npj Quantum Information},
  volume  = {11},
  pages   = {9},
  year    = {2025},
  doi     = {10.1038/s41534-025-00959-5}
}

@article{Li2025Ergotropy,
  author  = {Li, Li and Zhao, Silu and Shi, Yun-Hao and Xu, Kai and Fan, Heng and Zheng, Dongning and Xiang, Zhongcheng},
  title   = {Experimental extraction of coherent ergotropy and its energetic cost in a superconducting qubit},
  journal = {Physical Review Research},
  volume  = {7},
  number  = {4},
  pages   = {043236},
  year    = {2025},
  doi     = {10.1103/ptr1-59tv}
}

@article{Sandik2025,
  author  = {Sandik, Gal and Feist, Johannes and Garc{\'i}a-Vidal, Francisco J. and Schwartz, Tal},
  title   = {Cavity-enhanced energy transport in molecular systems},
  journal = {Nature Materials},
  volume  = {24},
  pages   = {344--355},
  year    = {2025},
  doi     = {10.1038/s41563-024-01962-5}
}

@article{Hu2024PolaritonTransfer,
  author  = {Hu, Zehua and Krisnanda, Tanjung and Fieramosca, Antonio and Zhao, Jiaxin and Sun, Qianlu and Chen, Yuzhong and Liu, Haiyun and Luo, Yuan and Su, Rui and Wang, Junyong and Watanabe, Kenji and Taniguchi, Takashi and Eda, Goki and Wang, Xiao Renshaw and Ghosh, Sanjib and Dini, Kevin and Sanvitto, Daniele and Liew, Timothy C. H. and Xiong, Qihua},
  title   = {Energy transfer driven brightening of {MoS$_2$} by ultrafast polariton relaxation in microcavity {MoS$_2$/hBN/WS$_2$} heterostructures},
  journal = {Nature Communications},
  volume  = {15},
  pages   = {1747},
  year    = {2024},
  doi     = {10.1038/s41467-024-45554-y}
}

@article{Lohse2025,
  author  = {Lohse, Leon M. and Andreji{\'c}, Petar and Velten, Sven and Vassholz, Malte and Neuhaus, Charlotte and Negi, Ankita and Panchwanee, Anjali and Sergeev, Ilya and P{\'a}lffy, Adriana and Salditt, Tim and R{\"o}hlsberger, Ralf},
  title   = {Collective Nuclear Excitation and Pulse Propagation in Single-Mode {X}-Ray Waveguides},
  journal = {Physical Review Letters},
  volume  = {135},
  number  = {5},
  pages   = {053601},
  year    = {2025},
  doi     = {10.1103/r2hf-9qn9}
}

@article{Lentrodt2025,
  author  = {Lentrodt, Dominik and Keitel, Christoph H. and Evers, J{\"o}rg},
  title   = {Toward Nonlinear Optics with {M{\"o}ssbauer} Nuclei Using {X}-Ray Cavities},
  journal = {Physical Review Letters},
  volume  = {135},
  number  = {3},
  pages   = {033801},
  year    = {2025},
  doi     = {10.1103/1lcz-r8d3}
}

@article{Brawley2025,
  author  = {Brawley, Zachary T. and Pannir-Sivajothi, Sindhana and Yim, Ju Eun and Poh, Yong Rui and Yuen-Zhou, Joel and Sheldon, Matthew},
  title   = {Vibrational weak and strong coupling modify a chemical reaction via cavity-mediated radiative energy transfer},
  journal = {Nature Chemistry},
  volume  = {17},
  pages   = {439--447},
  year    = {2025},
  doi     = {10.1038/s41557-024-01723-6}
}

@article{Muller2025,
  author  = {Muller, Cyprien and Piejko, Maciej and Bascil, Sinan and Moran, Joseph},
  title   = {The {Diels--Alder} Reaction as a Mechanistic Probe for Vibrational Strong Coupling},
  journal = {Angewandte Chemie International Edition},
  volume  = {64},
  number  = {36},
  pages   = {e202509391},
  year    = {2025},
  doi     = {10.1002/anie.202509391}
}

@article{Wang2024NFA,
  author  = {Wang, Jiayu and Xie, Yuan and Chen, Kai and Wu, Hongbin
             and Hodgkiss, Justin M. and Zhan, Xiaowei},
  title   = {Physical insights into non-fullerene organic photovoltaics},
  journal = {Nature Reviews Physics},
  volume  = {6},
  pages   = {365--381},
  year    = {2024},
  doi     = {10.1038/s42254-024-00719-y}
}

@article{Guo2024ExcitonDynamics,
  author  = {Guo, Y. and Han, G. and Guo, J. and Guo, H. and Fu, Y.
             and Miao, X. and Wang, Z. and Li, D. and Li, S. and Xu, X.
             and Lu, X. and Chen, H. and Yi, Y. and Chow, P. C. Y.},
  title   = {Engineering ultrafast exciton dynamics to boost organic
             photovoltaic performance},
  journal = {Energy \& Environmental Science},
  volume  = {17},
  pages   = {8776--8786},
  year    = {2024},
  doi     = {10.1039/D4EE03315A}
}

@article{Ivanovic2025,
  author  = {Ivanovi{\'c}, Filip and Giannini, Samuele and Peng, Wei-Tao
             and Blumberger, Jochen},
  title   = {Transiently delocalised hybrid quantum states are gateways
             for efficient exciton dissociation at organic donor-acceptor
             interfaces},
  journal = {Nature Communications},
  volume  = {16},
  pages   = {11560},
  year    = {2025},
  doi     = {10.1038/s41467-025-67722-4}
}

@article{Husel2026Dispersive,
  author  = {Husel, Lukas and Tabataba-Vakili, Farsane and Scherzer, Johannes
             and Krelle, Lukas and Bilgin, Ismail and Vadia, Samarth
             and Watanabe, Kenji and Taniguchi, Takashi and Carusotto, Iacopo
             and H{\"o}gele, Alexander},
  title   = {Cavity-mediated exciton hopping in a dielectrically engineered
             polariton system},
  journal = {Nature Communications},
  volume  = {17},
  pages   = {3779},
  year    = {2026},
  doi     = {10.1038/s41467-026-72043-1}
}

@article{Dijkema2025iSWAP,
  author  = {Dijkema, Jurgen and Xue, Xiao and Harvey-Collard, Patrick
             and Rimbach-Russ, Maximilian and de Snoo, Sander L.
             and Zheng, Guoji and Sammak, Amir and Scappucci, Giordano
             and Vandersypen, Lieven M. K.},
  title   = {Cavity-mediated {iSWAP} oscillations between distant spins},
  journal = {Nature Physics},
  volume  = {21},
  pages   = {168--174},
  year    = {2025},
  doi     = {10.1038/s41567-024-02694-8}
}

@article{Kertzscher2026PolaritonTransfer,
  author  = {Kertzscher, Christoph and Mauch, Michael and Keck, Jakob
             and Meixner, Alfred J.},
  title   = {Polariton-Mediated Energy Transfer in a Tunable Optical
             Multimode Microcavity Exceeding 10~{$\mu$}m},
  journal = {Advanced Photonics Research},
  volume  = {7},
  number  = {1},
  pages   = {e202500296},
  year    = {2026},
  doi     = {10.1002/adpr.202500296}
}

@article{Cao2022GeneralizedRET,
  author  = {Cao, Jianshu},
  title   = {Generalized Resonance Energy Transfer Theory: Applications
             to Vibrational Energy Flow in Optical Cavities},
  journal = {The Journal of Physical Chemistry Letters},
  volume  = {13},
  number  = {47},
  pages   = {10943--10951},
  year    = {2022},
  doi     = {10.1021/acs.jpclett.2c02707}
}

@article{Kushwaha2025Supertransfer,
  author  = {Kushwaha, Adesh and Kassal, Ivan},
  title   = {Engineering Quantum-Enhanced Transport by Supertransfer},
  journal = {arXiv preprint arXiv:2506.05045},
  year    = {2025},
  eprint  = {2506.05045},
  archivePrefix = {arXiv},
  primaryClass  = {quant-ph},
  doi     = {10.48550/arXiv.2506.05045}
}

@article{Green2026EfficiencyTables,
  author  = {Green, Martin A. and Dunlop, Ewan D. and Yoshita, Masahiro
             and Kopidakis, Nikos and Bothe, Karsten and Siefer, Gerald
             and Jiang, Jessica Yajie},
  title   = {Solar cell efficiency tables: Version 68},
  journal = {Joule},
  volume  = {10},
  number  = {7},
  pages   = {102494},
  year    = {2026},
  doi     = {10.1016/j.joule.2026.102494}
}

@article{Kang2009Ultrafast,
  author  = {Kang, Byoungwoo and Ceder, Gerbrand},
  title   = {Battery materials for ultrafast charging and discharging},
  journal = {Nature},
  volume  = {458},
  pages   = {190--193},
  year    = {2009},
  doi     = {10.1038/nature07853}
}

@book{Galison1997,
  author    = {Galison, Peter},
  title     = {Image and Logic: A Material Culture of Microphysics},
  publisher = {University of Chicago Press},
  address   = {Chicago},
  year      = {1997}
}

@book{Kaiser2005,
  author    = {Kaiser, David},
  title     = {Drawing Theories Apart: The Dispersion of Feynman Diagrams
               in Postwar Physics},
  publisher = {University of Chicago Press},
  address   = {Chicago},
  year      = {2005}
}

@article{DeLosh1970,
  author  = {DeLosh, Robert G. and Grant, Walter J. C.},
  title   = {Electronic Energy Transfer via Virtual-Phonon Processes},
  journal = {Physical Review B},
  volume  = {1},
  pages   = {1754--1760},
  year    = {1970},
  doi     = {10.1103/PhysRevB.1.1754}
}

@article{NagibarovaNagibarov1969,
  author  = {Nagibarova, I. A. and Nagibarov, V. R.},
  title   = {Exchange-Phonon Mechanism of Energy Migration in Solids},
  journal = {Soviet Physics JETP},
  volume  = {28},
  number  = {4},
  pages   = {669--674},
  year    = {1969}
}

@article{Blais2007CircuitQED,
  author  = {Blais, Alexandre and Gambetta, Jay and Wallraff, A.
             and Schuster, D. I. and Girvin, S. M. and Devoret, M. H.
             and Schoelkopf, R. J.},
  title   = {Quantum-information processing with circuit quantum
             electrodynamics},
  journal = {Physical Review A},
  volume  = {75},
  number  = {3},
  pages   = {032329},
  year    = {2007},
  doi     = {10.1103/PhysRevA.75.032329},
  eprint  = {cond-mat/0612038},
  archivePrefix = {arXiv}
}

@article{Allcock2000LARET,
  author  = {Allcock, Philip and Jenkins, Robert D. and Andrews, David L.},
  title   = {Laser-assisted resonance-energy transfer},
  journal = {Physical Review A},
  volume  = {61},
  number  = {2},
  pages   = {023812},
  year    = {2000},
  doi     = {10.1103/PhysRevA.61.023812}
}

@article{Caruso2009NoiseAssisted,
  author  = {Caruso, Filippo and Chin, Alex W. and Datta, Animesh
             and Huelga, Susana F. and Plenio, Martin B.},
  title   = {Highly efficient energy excitation transfer in
             light-harvesting complexes: The fundamental role of
             noise-assisted transport},
  journal = {The Journal of Chemical Physics},
  volume  = {131},
  number  = {10},
  pages   = {105106},
  year    = {2009},
  doi     = {10.1063/1.3223548}
}

@article{Potocnik2018LightHarvesting,
  author  = {Poto{\v{c}}nik, Anton and Bargerbos, Arno
             and Schr{\"o}der, Florian A. Y. N. and Khan, Saeed A.
             and Collodo, Michele C. and Gasparinetti, Simone
             and Salath{\'e}, Yves and Creatore, Celestino
             and Eichler, Christopher and T{\"u}reci, Hakan E.
             and Chin, Alex W. and Wallraff, Andreas},
  title   = {Studying light-harvesting models with superconducting circuits},
  journal = {Nature Communications},
  volume  = {9},
  number  = {1},
  pages   = {904},
  year    = {2018},
  doi     = {10.1038/s41467-018-03312-x}
}

@article{Viciani2015NoiseAssisted,
  author  = {Viciani, Silvia and Lima, Manuela and Bellini, Marco
             and Caruso, Filippo},
  title   = {Observation of Noise-Assisted Transport in an
             All-Optical Cavity-Based Network},
  journal = {Physical Review Letters},
  volume  = {115},
  number  = {8},
  pages   = {083601},
  year    = {2015},
  doi     = {10.1103/PhysRevLett.115.083601}
}

@article{Dovzhenko2021RoleReversal,
  author  = {Dovzhenko, Dmitriy and Lednev, Maksim and Mochalov, Konstantin
             and Vaskan, Ivan and Rakovich, Yury and Karaulov, Alexander
             and Nabiev, Igor},
  title   = {Polariton-assisted manipulation of energy relaxation pathways:
             donor--acceptor role reversal in a tuneable microcavity},
  journal = {Chemical Science},
  volume  = {12},
  pages   = {12794--12805},
  year    = {2021},
  doi     = {10.1039/D1SC02026A}
}

@article{Polak2020TripletHarvesting,
  author  = {Polak, Daniel and Jayaprakash, Rahul and Lyons, Thomas P.
             and Mart{\'i}nez-Mart{\'i}nez, Luis {\'A}.
             and Leventis, Anastasia and Fallon, Kealan J.
             and Coulthard, Harriet and Bossanyi, David G.
             and Georgiou, Kyriacos and Petty, II, Anthony J.
             and Anthony, John and Bronstein, Hugo and Yuen-Zhou, Joel
             and Tartakovskii, Alexander I. and Clark, Jenny
             and Musser, Andrew J.},
  title   = {Manipulating molecules with strong coupling: harvesting triplet
             excitons in organic exciton microcavities},
  journal = {Chemical Science},
  volume  = {11},
  pages   = {343--354},
  year    = {2020},
  doi     = {10.1039/C9SC04950A}
}

@article{Marsault2015PolaritonRouter,
  author  = {Marsault, F{\'e}lix and Nguyen, Hai Son and Tanese, Dimitrii
             and Lema{\^i}tre, Aristide and Galopin, Elisabeth and Sagnes, Isabelle
             and Amo, Alberto and Bloch, Jacqueline},
  title   = {Realization of an all optical exciton-polariton router},
  journal = {Applied Physics Letters},
  volume  = {107},
  number  = {20},
  pages   = {201115},
  year    = {2015},
  doi     = {10.1063/1.4936158}
}

@article{Liu2020ExcitonRouter,
  author  = {Liu, Yuanda and Dini, K{\'e}vin and Tan, Qinghai
             and Liew, Timothy and Novoselov, Kostya S. and Gao, Weibo},
  title   = {Electrically controllable router of interlayer excitons},
  journal = {Science Advances},
  volume  = {6},
  number  = {41},
  pages   = {eaba1830},
  year    = {2020},
  doi     = {10.1126/sciadv.aba1830}
}

@article{Miao2025QuantumRouter,
  author  = {Miao, Connie and L{\'e}ger, S{\'e}bastien and Li, Ziqian
             and Lee, Gideon and Jiang, Liang and Schuster, David I.},
  title   = {Implementation of a Quantum Addressable Router Using
             Superconducting Qubits},
  journal = {PRX Quantum},
  volume  = {6},
  number  = {4},
  pages   = {040335},
  year    = {2025},
  doi     = {10.1103/pq3x-cmw9}
}

@article{Kersten2023Triggered,
  author = {Kersten, Wenzel and de Zordo, Nikolaus and Diekmann, Oliver
            and Reiter, Tobias and Zens, Matthias and Kanagin, Andrew N.
            and Rotter, Stefan and Schmiedmayer, J{\"o}rg and Angerer, Andreas},
  title = {Triggered Superradiance and Spin Inversion Storage in a Hybrid Quantum System},
  journal = {Physical Review Letters},
  volume = {131},
  number = {4},
  pages = {043601},
  year = {2023},
  doi = {10.1103/PhysRevLett.131.043601},
  eprint = {2301.04100},
  archivePrefix = {arXiv},
  url = {https://arxiv.org/abs/2301.04100v2}
}

@article{Straight2005DoubleThrow,
  author  = {Straight, Stephen D. and Andr{\'e}asson, Joakim
             and Kodis, Gerdenis and Moore, Ana L.
             and Moore, Thomas A. and Gust, Devens},
  title   = {Photochromic Control of Photoinduced Electron Transfer.
             Molecular Double-Throw Switch},
  journal = {Journal of the American Chemical Society},
  volume  = {127},
  number  = {8},
  pages   = {2717--2724},
  year    = {2005},
  doi     = {10.1021/ja044128i}
}

@article{Shvydko1996,
  author  = {Shvyd'ko, Yu. V. and Hertrich, T. and van B{\"u}rck, U.
             and Gerdau, E. and Leupold, O. and Metge, J.
             and R{\"u}ter, H. D. and Schwendy, S. and Smirnov, G. V.
             and Potzel, W. and Schindelmann, P.},
  title   = {Storage of Nuclear Excitation Energy through Magnetic Switching},
  journal = {Physical Review Letters},
  volume  = {77},
  number  = {15},
  pages   = {3232--3235},
  year    = {1996},
  doi     = {10.1103/PhysRevLett.77.3232}
}

@article{Nazeeri2026SAW,
  author  = {Nazeeri, Albert and Brandenstein, Chiara and Jia, Chengjie
             and Magrini, Lorenzo and Gratta, Giorgio},
  title   = {Coupling of a Nuclear Transition to a Surface Acoustic Wave},
  journal = {Physical Review Letters},
  volume  = {136},
  number  = {18},
  pages   = {183801},
  year    = {2026},
  doi     = {10.1103/tc97-98f7},
  eprint  = {2509.02750},
  archivePrefix = {arXiv},
  primaryClass  = {quant-ph}
}

@article{Kishimoto2000NEET,
  author  = {Kishimoto, S. and Yoda, Y. and Seto, M. and Kobayashi, Y.
             and Kitao, S. and Haruki, R. and Kawauchi, T.
             and Fukutani, K. and Okano, T.},
  title   = {Observation of Nuclear Excitation by Electron Transition
             in {$^{197}$Au} with Synchrotron {X} Rays and an
             Avalanche Photodiode},
  journal = {Physical Review Letters},
  volume  = {85},
  number  = {9},
  pages   = {1831--1834},
  year    = {2000},
  doi     = {10.1103/PhysRevLett.85.1831}
}

@article{Kekez1985ElectronicBridge,
  author  = {Kekez, D. and Ljubicic, A. and Pisk, K. and Logan, B. A.},
  title   = {Nuclear Deexcitation via the Electronic-Bridge Mechanism},
  journal = {Physical Review Letters},
  volume  = {55},
  number  = {13},
  pages   = {1366--1368},
  year    = {1985},
  doi     = {10.1103/PhysRevLett.55.1366}
}

@article{Chumakov2018XFEL,
  author  = {Chumakov, Aleksandr I. and Baron, Alfred Q. R. and Sergueev, Ilya
             and Strohm, Cornelius and Leupold, Olaf and Shvyd'ko, Yuri
             and Smirnov, Gennadi V. and R{\"u}ffer, Rudolf
             and Inubushi, Yuichi and Yabashi, Makina and Tono, Kensuke
             and Kudo, Togo and Ishikawa, Tetsuya},
  title   = {Superradiance of an Ensemble of Nuclei Excited by a Free Electron Laser},
  journal = {Nature Physics},
  volume  = {14},
  pages   = {261--264},
  year    = {2018},
  doi     = {10.1038/s41567-017-0001-z}
}

@article{Loher2013NRF,
  author  = {L{\"o}her, Bastian and Derya, Vera and Aumann, Thomas
             and Beller, Jacob and Cooper, Nathan and Duch{\^e}ne, Marc
             and Endres, Janis and Fiori, Enrico and Isaak, Johann
             and Kelley, John H. and Kn{\"o}rzer, Michael
             and Pietralla, Norbert and Romig, Christopher
             and Savran, Deniz and Scheck, Marcus and Scheit, Heiko
             and Silva, Joel and Tonchev, Anton P. and Tornow, Werner
             and Weller, Henry R. and Werner, Volker and Zilges, Andreas},
  title   = {The High-Efficiency $\gamma$-Ray Spectroscopy Setup $\gamma^3$ at {HI$\gamma$S}},
  journal = {Nuclear Instruments and Methods in Physics Research Section A:
             Accelerators, Spectrometers, Detectors and Associated Equipment},
  volume  = {723},
  pages   = {136--142},
  year    = {2013},
  doi     = {10.1016/j.nima.2013.04.087},
  eprint  = {1304.6270},
  archivePrefix = {arXiv},
  primaryClass  = {physics.ins-det}
}

@article{WalkerDracoulis1999,
  author  = {Walker, Philip M. and Dracoulis, George D.},
  title   = {Energy traps in atomic nuclei},
  journal = {Nature},
  volume  = {399},
  number  = {6731},
  pages   = {35--40},
  year    = {1999},
  doi     = {10.1038/19911}
}

@article{WalkerPodolyak2020,
  author  = {Walker, Philip M. and Podoly{\'a}k, Zsolt},
  title   = {100 years of nuclear isomers---then and now},
  journal = {Physica Scripta},
  volume  = {95},
  number  = {4},
  pages   = {044004},
  year    = {2020},
  doi     = {10.1088/1402-4896/ab635d}
}

@article{Carroll2007,
  author  = {Carroll, James J.},
  title   = {Nuclear structure and the search for induced energy release from isomers},
  journal = {Nuclear Instruments and Methods in Physics Research Section B:
             Beam Interactions with Materials and Atoms},
  volume  = {261},
  number  = {1--2},
  pages   = {960--964},
  year    = {2007},
  doi     = {10.1016/j.nimb.2007.04.128}
}

@article{Rohlsberger2010Superradiance,
  author  = {R{\"o}hlsberger, Ralf and Schlage, Kai and Sahoo, Balaram
             and Couet, Sebastien and R{\"u}ffer, Rudolf},
  title   = {Collective Lamb Shift in Single-Photon Superradiance},
  journal = {Science},
  volume  = {328},
  number  = {5983},
  pages   = {1248--1251},
  year    = {2010},
  doi     = {10.1126/science.1187770}
}

@article{Belic2002Ta180m,
  author  = {Beli{\'c}, D. and Arlandini, C. and Besserer, J. and others},
  title   = {Photo-induced depopulation of the {$^{180\mathrm{m}}$Ta}
             isomer via low-lying intermediate states:
             Structure and astrophysical implications},
  journal = {Physical Review C},
  volume  = {65},
  pages   = {035801},
  year    = {2002},
  doi     = {10.1103/PhysRevC.65.035801}
}

@article{Ding2026Mo93m,
  author  = {Ding, B. and Jia, C. X. and Guo, S. and others},
  title   = {Isomer Depletion of {$^{93\mathrm{m}}$Mo} Triggered by
             Inelastic Nuclear Scattering Rather than Nuclear Excitation
             by Electron Capture},
  journal = {Physical Review Letters},
  volume  = {136},
  pages   = {052502},
  year    = {2026},
  doi     = {10.1103/kbf5-6fcl}
}

@article{Huke2006,
  author  = {Huke, A. and Czerski, K. and Dorsch, T. and Biller, A.
             and Heide, P. and Ruprecht, G.},
  title   = {Evidence for a host-material dependence of the n/p
             branching ratio of low-energy d+d reactions within
             metallic environments},
  journal = {The European Physical Journal A},
  volume  = {27},
  pages   = {187--192},
  year    = {2006},
  doi     = {10.1140/epja/i2006-08-028-3}
}

@article{Dubey2025,
  author  = {Dubey, R. and Czerski, K. and Das H., Gokul and Kowalska, A.
             and Targosz-{\'S}leczka, N. and Kaczmarski, M. and Valat, M.},
  title   = {Experimental Signatures of a New Channel of the
             Deuteron-Deuteron Reaction at Very Low Energy},
  journal = {Physical Review X},
  volume  = {15},
  pages   = {041004},
  year    = {2025},
  doi     = {10.1103/chlp-b215}
}

@article{Paetz1992PolarizedDD,
  author  = {Paetz gen. Schieck, H. and Becker, B. and Randermann, R.
             and Lema{\^i}tre, S. and Niessen, P.
             and Reckenfelderb{\"a}umer, R. and Sydow, L.},
  title   = {Non-suppression of quintet states in the tensor analyzing
             powers of the D+D fusion reactions at
             {$E_{\mathrm{cm}}=14$ keV}},
  journal = {Physics Letters B},
  volume  = {276},
  pages   = {290--293},
  year    = {1992},
  doi     = {10.1016/0370-2693(92)90320-4}
}

@article{Shvydko2023Scandium,
  author  = {Shvyd'ko, Yuri and R{\"o}hlsberger, Ralf
             and Kocharovskaya, Olga and Evers, J{\"o}rg
             and Geloni, Gianluca Aldo and Liu, Peifan and Shu, Deming
             and Miceli, Antonino and Stone, Brandon and Hippler, Willi
             and Marx-Glowna, Berit and Uschmann, Ingo and Loetzsch, Robert
             and Leupold, Olaf and Wille, Hans-Christian and Sergeev, Ilya
             and Gerharz, Miriam and Zhang, Xiwen and Grech, Christian
             and Guetg, Marc and Kocharyan, Vitali and Kujala, Naresh
             and Liu, Shan and Qin, Weilun and Zozulya, Alexey
             and Hallmann, J{\"o}rg and Boesenberg, Ulrike and Jo, Wonhyuk
             and M{\"o}ller, Johannes and Rodriguez-Fernandez, Angel
             and Youssef, Mohamed and Madsen, Anders and Kolodziej, Tomasz},
  title   = {Resonant {X}-ray excitation of the nuclear clock isomer {$^{45}$Sc}},
  journal = {Nature},
  volume  = {622},
  pages   = {471--475},
  year    = {2023},
  doi     = {10.1038/s41586-023-06491-w}
}

@article{Whyte2026Economics,
  author  = {Whyte, D. G. and Lo, A. and Bielajew, R.
             and Hancock, M. and Moeykens, R. and Shaw, G.},
  title   = {Criteria for the economic viability of fusion power plants},
  journal = {Journal of Fusion Energy},
  volume  = {45},
  pages   = {49},
  year    = {2026},
  doi     = {10.1007/s10894-026-00577-9}
}

@article{Tang2024NeutronStates,
  author  = {Tang, Hao and Wang, Guoqing and Cappellaro, Paola
             and Li, Ju},
  title   = {{$\mu$eV}-Deep Neutron Bound States in Nanocrystals},
  journal = {ACS Nano},
  volume  = {18},
  number  = {12},
  pages   = {9063--9070},
  year    = {2024},
  doi     = {10.1021/acsnano.3c12929},
  url     = {https://doi.org/10.1021/acsnano.3c12929}
}

@article{Tiedau2024,
  author  = {Tiedau, J. and Okhapkin, M. V. and Zhang, K.
             and Thielking, J. and Zitzer, G. and Peik, E.
             and Schaden, F. and Pronebner, T. and Morawetz, I.
             and De Col, L. Toscani and Schneider, F. and Leitner, A.
             and Pressler, M. and Kazakov, G. A. and Beeks, K.
             and Sikorsky, T. and Schumm, T.},
  title   = {Laser Excitation of the {Th-229} Nucleus},
  journal = {Physical Review Letters},
  volume  = {132},
  number  = {18},
  pages   = {182501},
  year    = {2024},
  doi     = {10.1103/PhysRevLett.132.182501}
}

@article{Elwell2024,
  author  = {Elwell, R. and Schneider, Christian and Jeet, Justin
             and Terhune, J. E. S. and Morgan, H. W. T.
             and Alexandrova, A. N. and Tran Tan, H. B.
             and Derevianko, Andrei and Hudson, Eric R.},
  title   = {Laser Excitation of the {$^{229}$Th} Nuclear Isomeric
             Transition in a Solid-State Host},
  journal = {Physical Review Letters},
  volume  = {133},
  number  = {1},
  pages   = {013201},
  year    = {2024},
  doi     = {10.1103/PhysRevLett.133.013201}
}

@article{Thomas2019Selectivity,
  author  = {Thomas, Anoop and Lethuillier-Karl, Lucas
             and Nagarajan, Kalaivanan and Vergauwe, Robrecht M. A.
             and George, J. and Chervy, T.
             and Shalabney, A. and Devaux, E.
             and Genet, C. and Moran, J. and Ebbesen, T. W.},
  title   = {Tilting a ground-state reactivity landscape by
             vibrational strong coupling},
  journal = {Science},
  volume  = {363},
  number  = {6427},
  pages   = {615--619},
  year    = {2019},
  doi     = {10.1126/science.aau7742}
}

@article{Sau2021Stereoselectivity,
  author  = {Sau, Abhijit and Nagarajan, Kalaivanan
             and Patrahau, Bianca and Lethuillier-Karl, Lucas
             and Vergauwe, Robrecht M. A. and Thomas, Anoop
             and Moran, Joseph and Genet, Cyriaque
             and Ebbesen, Thomas W.},
  title   = {Modifying {Woodward--Hoffmann} Stereoselectivity
             Under Vibrational Strong Coupling},
  journal = {Angewandte Chemie International Edition},
  volume  = {60},
  number  = {11},
  pages   = {5712--5717},
  year    = {2021},
  doi     = {10.1002/anie.202013465}
}

@misc{DARPADecayOnDemand,
  author       = {{Defense Advanced Research Projects Agency}},
  title        = {{Decay on Demand}},
  howpublished = {Program description, Defense Sciences Office},
  url          = {https://www.darpa.mil/research/programs/decay-on-demand},
  note         = {Accessed 22 September 2026}
}

@article{Deltuva2010PolarizedDD,
  author  = {Deltuva, A. and Fonseca, A. C.},
  title   = {Polarization observables and spin-aligned fusion rates in
             {$^{2}\mathrm{H}(d,p)^{3}\mathrm{H}$} and
             {$^{2}\mathrm{H}(d,n)^{3}\mathrm{He}$} reactions},
  journal = {Physical Review C},
  volume  = {81},
  number  = {5},
  pages   = {054002},
  year    = {2010},
  doi     = {10.1103/PhysRevC.81.054002}
}

@article{Viviani2023DD,
  author  = {Viviani, M. and Girlanda, L. and Kievsky, A. and
             Logoteta, D. and Marcucci, L. E.},
  title   = {Theoretical Study of the
             {$d(d,p)^{3}\mathrm{H}$} and
             {$d(d,n)^{3}\mathrm{He}$} Processes at Low Energies},
  journal = {Physical Review Letters},
  volume  = {130},
  number  = {12},
  pages   = {122501},
  year    = {2023},
  doi     = {10.1103/PhysRevLett.130.122501}
}

@article{Hagelstein2008ExcitationTransfer,
  author  = {Hagelstein, P. L. and Chaudhary, I. U.},
  title   = {Excitation transfer in two two-level systems coupled to an oscillator},
  journal = {Journal of Physics B: Atomic, Molecular and Optical Physics},
  volume  = {41},
  number  = {13},
  pages   = {135501},
  year    = {2008},
  doi     = {10.1088/0953-4075/41/13/135501}
}

@article{Jones2022NeutronTransfer,
  author  = {Jones, K. L. and Bey, A. and Burcher, S. and Allmond, J. M. and
             Galindo-Uribarri, A. and Radford, D. C. and Ahn, S. and
             Ayres, A. and Bardayan, D. W. and Cizewski, J. A. and others},
  title   = {Neutron transfer reactions on the ground state and isomeric state
             of a {$^{130}$Sn} beam},
  journal = {Physical Review C},
  volume  = {105},
  number  = {2},
  pages   = {024602},
  year    = {2022},
  doi     = {10.1103/PhysRevC.105.024602}
}

@article{Timofeyuk2020Transfer,
  author  = {Timofeyuk, N. K. and Johnson, R. C.},
  title   = {Theory of deuteron stripping and pick-up reactions for nuclear
             structure studies},
  journal = {Progress in Particle and Nuclear Physics},
  volume  = {111},
  pages   = {103738},
  year    = {2020},
  doi     = {10.1016/j.ppnp.2019.103738}
}

@article{AsenjoGarcia2017SelectiveRadiance,
  author  = {Asenjo-Garcia, A. and Moreno-Cardoner, M.
             and Albrecht, A. and Kimble, H. J. and Chang, D. E.},
  title   = {Exponential Improvement in Photon Storage Fidelities
             Using Subradiance and ``Selective Radiance''
             in Atomic Arrays},
  journal = {Physical Review X},
  volume  = {7},
  number  = {3},
  pages   = {031024},
  year    = {2017},
  doi     = {10.1103/PhysRevX.7.031024}
}

@article{Sharma2020Supertransport,
  author  = {Sharma, Ankur and Zhang, Linglong and
             Tollerud, Jonathan O. and Dong, Miheng and
             Zhu, Yi and Halbich, Robert and Vogl, Tobias and
             Liang, Kun and Nguyen, Hieu T. and Wang, Fan and
             Sanwlani, Shilpa and Earl, Stuart K. and
             Macdonald, Daniel and Lam, Ping Koy and
             Davis, Jeffrey A. and Lu, Yuerui},
  title   = {Supertransport of excitons in atomically thin organic
             semiconductors at the {2D} quantum limit},
  journal = {Light: Science \& Applications},
  volume  = {9},
  pages   = {116},
  year    = {2020},
  doi     = {10.1038/s41377-020-00347-y}
}

\end{document}